\documentclass[a4paper,fleqn]{cas-dc}
\usepackage{algorithm}
\usepackage{algorithmic}
\usepackage[authoryear]{natbib}
\usepackage{hyperref}
\usepackage{color,soul}
\usepackage{booktabs}
\usepackage{graphicx}
\usepackage{adjustbox}
\usepackage{array}
\usepackage{amsmath,amsfonts}
\usepackage{array} 
\usepackage{fvextra}
\usepackage{enumitem}
\DefineVerbatimEnvironment{MyVerbatim}{Verbatim}{breaklines=true}
\usepackage{subcaption} 

\def\tsc#1{\csdef{#1}{\textsc{\lowercase{#1}}\xspace}}
\tsc{WGM}
\tsc{QE}
\tsc{EP}
\tsc{PMS}
\tsc{BEC}
\tsc{DE}

\usepackage{xcolor}

\usepackage{pythonhighlight}

\begin{document}
\let\WriteBookmarks\relax
\def\floatpagepagefraction{1}
\def\textpagefraction{.001}

\shorttitle{Towards a Barrier-free GeoQA Portal}

\shortauthors{Feng et~al.}

\title [mode = title]{Towards a Barrier-free GeoQA Portal: Natural Language Interaction with Geospatial Data Using Multi-Agent LLMs and Semantic Search}  
              
%
\author[1]{Yu Feng}[orcid=0000-0001-5110-5564]


\ead{y.feng@tum.de}


\credit{Writing - Original Draft, Writing - Review \& Editing, Conceptualization, Methodology, Validation, Supervision, Project administration, Funding acquisition}

\author[1]{Puzhen Zhang}[orcid=0009-0004-2208-0942]

\cormark[1]


\ead{puzhen.zhang@tum.de}

\credit{Methodology, Validation, Software, Data Curation, Writing - Original Draft, Writing - Review \& Editing, Formal analysis, Visualization}

\affiliation[1]{organization={Chair of Cartography and Visual Analytics, Technical University of Munich},
    addressline={Arcisstrasse 21}, 
    city={Munich},
    postcode={80333},
    country={Germany}}

\author[2]{Guohui Xiao}[orcid=0000-0002-5115-4769]

\ead{guohui.xiao@seu.edu.cn}

\affiliation[2]{organization={School of Computer Science and Engineering, Southeast University},
    city={Nanjing},
    postcode={210096},
    country={China}}


\credit{Writing - Original Draft, Writing - Review \& Editing, Supervision, Funding acquisition}

\author[3]{Linfang Ding}[orcid=0000-0002-3707-5845]

\ead{linfang.ding@hhu.edu.cn}

\affiliation[3]{organization={College of Geography and Remote Sensing, Hohai University},
    city={Nanjing},
    postcode={211000},
    country={China}}
    

\credit{Writing - Review \& Editing, Supervision, Funding acquisition}

\author[1]{Liqiu Meng}[orcid=0000-0001-8787-3418]

\ead{liqiu.meng@tum.de}

\credit{Writing - Original Draft, Writing - Review \& Editing, Project administration, Resources, Funding acquisition}

\cortext[cor1]{Corresponding author}

\begin{abstract}
Geoportals have become essential tools for accessing and analyzing geospatial data, facilitating open spatial data sharing and online geo-information management. 
They typically follow interaction logic similar to common GIS systems and use a layered visualization structure. 
However, these designs can make geoportals less accessible to non-expert users, who may find complex functionalities overwhelming and struggle with overlapping data layers that obscure spatial relationships.

To address these challenges, we propose a step towards a barrier-free GeoQA Portal that enables seamless natural language interaction with geospatial data using a multi-agent Large Language Model (LLM) framework.
By decomposing user queries into individual subtasks and assigning them to different LLM agents, our portal can process complex spatial queries and extract geographic data relevant to user interests. 
Additionally, task plans are presented to users, improving transparency and engagement. 
Our geoportal supports both default and custom data inputs, offering flexibility for diverse user needs.
Furthermore, to improve accessibility, we implement semantic search based on the similarity of word vector representation, allowing users to retrieve relevant data even if terms or spellings are imperfect.
Through case studies, system evaluation, and user testing, we demonstrate the system’s effectiveness in meeting the geospatial data access and analysis needs of non-expert users. 
The results highlight the ability of our portal to bridge the gap between complex GIS workflows and public data accessibility, offering a more intuitive and user-friendly solution for future geoportal development. 
\end{abstract}

\begin{highlights}
\item Designed and developed a geoportal for intuitive GIS data retrieval and analysis through natural language queries. 
\item Proposed a multi-agent-based framework that integrates transparent reasoning procedures with interactive visualizations.
\item Conducted systematic evaluation and user tests to demonstrate the high performance of this natural language-based geoportal.
\end{highlights}


\maketitle

\section{Introduction}

Over the past few decades, Geographic Information System (GIS) technology has made significant advancements. However, for non-expert users without specialized knowledge, querying and analyzing geospatial data remains a challenge. Using GIS software typically requires professional expertise and training. 
Many online geoportals offer data visualization and basic GIS tools aimed at lowering the barrier for public use. 

However, these systems typically follow an interaction logic similar to common GIS platforms and adopt a layered display structure. 
For non-expert users, overlaying multiple data layers often leads to information overload, making it difficult to locate specific areas of interest or perform targeted analysis. 
This complexity can easily cause users to lose direction during the data visualization process. 
Moreover, complex queries across data layers or databases often require users to fill out structured query forms or construct queries like SQL, which poses a significant barrier for users without a technical background.

As a result, Geographic Question Answering (GeoQA) has increasingly gained attention from researchers. 
Providing answers to non-expert users' questions in a Q\&A format could significantly enhance the usability of GIS and online geoportals. 
Geographic questions can be categorized into eight types: factual, predictive, opinion, hypothetical, causal, geo-analytical, scenario-based, and visual questions \citep{mai2021geographic}.
Among these, factual and geo-analytical questions are particularly unique and challenging, and are one of the most common types faced by general users. These questions often involve geoprocessing workflows and can have answers in various forms, including not only text but also geographic entities or maps. This highlights an exciting future for GIS, where geospatial retrieval and analysis processes could be automated without human intervention.

With the advances of Large Language Models (LLMs) over the past years, various innovative methods have been proposed to address geo-analytical questions, including GeoQAMap \citep{feng2023geoqamap}, Text-to-OverpassQL \citep{staniek2023texttooverpassql}, Autonomous GIS \citep{ning2024autonomous}, MapGPT \citep{zhang2024mapgpt}, GeoGPT \citep{zhang2024geogpt}, etc.
GeoQAMap and OverpassQL can directly access spatial databases or knowledge bases, but they rely heavily on LLM-based SPARQL or Overpass query generation. 
This reliance often leads to errors in queries, compromising the reliability and accuracy of the systems.
On the other hand, Autonomous GIS, MapGPT, and GeoGPT resemble traditional GIS tools, requiring users to explicitly specify data sources, which still presents a barrier for non-expert users. 
Furthermore, due to the generative nature of LLMs, the answers provided by these systems carry inherent uncertainty. As datasets become more complex and the questions more intricate, the risk of errors increases, making it  more challenging for these systems to effectively handle complex geographic problems \citep{zhang2024geogpt}.

More importantly, all of the above methods require the input of complete name spelling or concepts that exactly match the dataset, which is not realistic for non-expert users. As a result, users are likely to feel frustrated when they fail to obtain the correct match due to their inability to accurately specify the data. 

Such a scenario inspired us to develop a barrier-free geoportal that enables users to interact with geospatial data seamlessly through natural language.
The system utilizes a multi-agent LLM framework to provide greater robustness in handling variations in question phrasing while seamlessly accessing multiple types of data sources. 
Users can utilize the default data settings of the system or input custom external data, offering flexibility and improving accessibility for novice users and their specific needs.
By decomposing complex tasks and assigning them to different LLM agents, with each agent focusing on a specialized subtask, a similar approach has shown significant improvement in the overall success rate~\citep{wang2024tdag}.
The task-planning agent formulates detailed steps for each reasoning task and presents them to the user, ensuring that the reasoning process is transparent.
In addition, we specifically consider semantic search and utilize word vector representation similarity to enhance the flexibility and accuracy of matching.
Furthermore, similar existing tools have mainly demonstrated their potential without thoroughly verifying their alignment with user needs and experience. 
Therefore, our work placed additional emphasis on incorporating user feedback to ensure the tool’s utility.

The subsequent sections are structured as follows:
Section 2 provides a comprehensive overview of related works.
In Section 3, we introduce the workflow of the proposed framework.
Section 4 delves into the experiment and results, where we present the case studies, system performance evaluation, and user tests.
In Section 5, we provide detailed insights gained from the performance evaluation and user study.
Finally, in Section 6, we conclude and provide an outlook.




\section{Related Work}

This study focuses on natural language queries for databases. Therefore, we review relevant research in this area in Section \ref{sec:llm_query} and explore the latest advancements in LLMs within the field of GIS in Section \ref{sec:llm_gis}.

\subsection{Natural language querying databases with LLM}
\label{sec:llm_query}
In the context of accessing databases through natural language, advances in NLP have driven the evolution of Natural language interfaces to databases (NLIDB) from early template-based, rule-driven methods to more sophisticated machine learning and deep learning approaches.
Early systems \citep[e.g.,][]{10.1145/604045.604070,androutsopoulos1995natural} primarily relied on manually designed patterns to map natural language queries to SQL statements. 
These systems have limited flexibility and are not adept at handling complex queries. 
By the mid-2000s, grammar-based and dependency tree parsing methods \citep[e.g.,][]{10.3115/1220355.1220376} were introduced, improving query parsing accuracy, but still required significant manual rule design.
In the 2010s, machine learning methods \citep[e.g.,][]{cai-yates-2013-large} reduced reliance on handcrafted rules. Later, works such as \cite{zhong2017seq2sql} and \cite{xu2017sqlnet} introduced sequence-to-sequence models and attention mechanisms to automate natural language-to-SQL translation, though challenges with complex queries persisted.
In recent years, large-scale pretrained models like BERT \citep{devlin2019bertpretrainingdeepbidirectional} became dominant, with works such as \cite{yu2019spiderlargescalehumanlabeleddataset, wang2021ratsqlrelationawareschemaencoding, guo-etal-2019-towards, zhang-etal-2019-editing} using pretraining to enhance semantic understanding across domains. While these models improved accuracy, challenges like computational cost, data dependency, and interpretability persist.

LLMs have enhanced language understanding and generalization and can handle natural language database queries with less data.
However, based on a benchmark of 12,751 text-to-SQL pairs, GPT-4 achieves an execution accuracy of only 54.89\%, compared to 92.96\% for humans \citep{li2024can}. 
Directly generating query strings poses challenges for fuzzy searches, as SQL-based approaches rely on exact or partial matches through expressions.
\cite{kovriguina2023sparqlgen,pliukhin2023improving} introduces SPARQLGEN, a framework that uses LLMs to generate SPARQL queries by employing embedding-based vector similarity to identify relevant subgraphs in the knowledge base. This improves the accuracy of LLM-generated queries for knowledge-based querying using natural language.

\subsection{Recent advances in GIS domain} 
\label{sec:llm_gis}
The application of LLMs in the field of GIS has increasingly garnered the attention of many researchers. Currently, the mainstream approaches to solving geo-analytical problems can be broadly divided into two categories.

The first category primarily relies on the code generation capabilities of LLMs.
For instance, GeoQAMap \citep{feng2023geoqamap} generates GeoSPARQL queries related to geographic questions to access the Wikidata\footnote{Wikidata: \url{https://www.wikidata.org/wiki/Wikidata:Main_Page}} knowledge base and visualizes the results on a map. 
This system can answer simple factual geographic questions and geo-analytical questions (e.g., buffer analysis). 
However, the most common errors are mismatches between the entity names used by users and those in the knowledge base, mainly due to the ambiguity of language.
Additionally, the generated queries could fail to execute properly, requiring manual intervention and corrections. 
Another example is given by \cite{staniek2023texttooverpassql}, which uses GPT-4 to generate Overpass Query Language (OverpassQL) designed specifically for querying OpenStreetMap (OSM) data. The research found that GPT-4 outperforms some fine-tuned models, but its performance declines on more challenging test cases, with an execution accuracy (EX) of only 22.2\% on the "Hard Partition."

\begin{figure*}[t]
\centering
\includegraphics[width=1.10\textwidth]{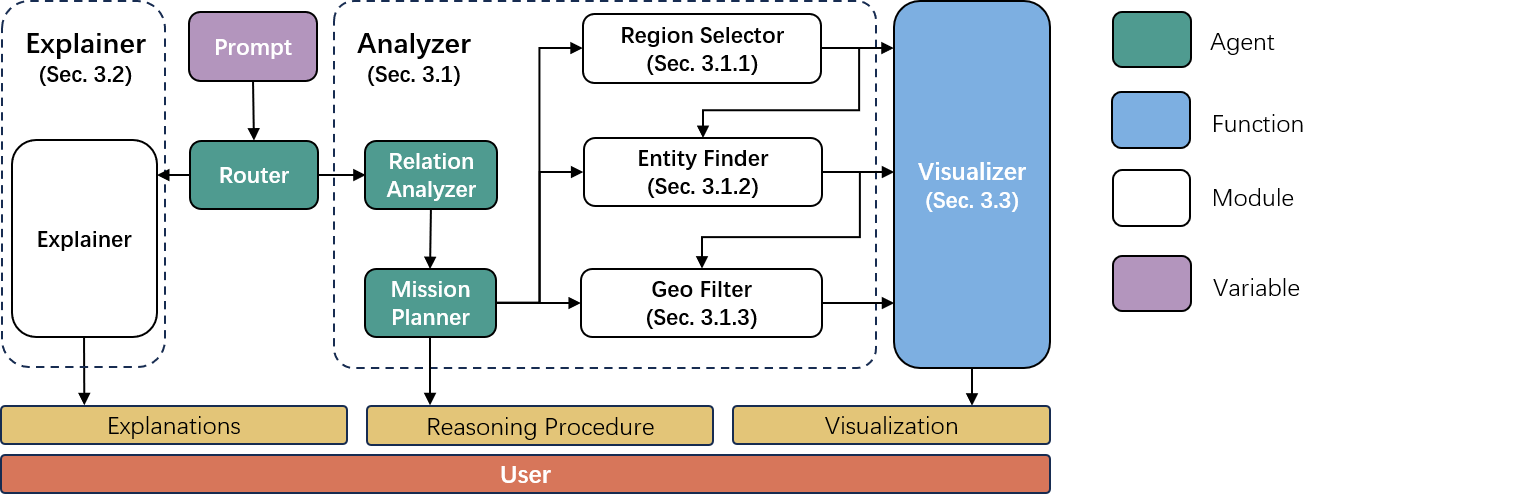}
\caption{Overview of the proposed framework. It contains three main components: (1) \textit{Analyzer}, (2) \textit{Explainer}, and (3) \textit{Visualizer}. Each component would include LLM agents or pre-implemented functions. The framework provides three possible outputs for users: explanations, reasoning procedures, and visualizations. Agents handle dynamic reasoning, while pre-implemented functions perform specific computational tasks. Different colors in the figure distinguish key processing units.}
\label{fig:framework}
\end{figure*}

Another category focuses on leveraging LLMs to break down complex geo-analytical problems into discrete, manageable steps and solve them progressively.
Autonomous GIS \citep{li2023autonomous} introduced a novel idea of using LLM generating solution graphs, to decompose the task into individual steps. 
Python code are generated for each step and then assembled to a fully executable program. 
However, it was mentioned that the generated code may not run directly and could contain syntax errors, API call errors, or logical errors. Furthermore, in real-world applications, GIS data is often dispersed across various sources. Users are typically required to manually download this data, specify file paths, or provide API endpoints to access it. 
Its further development of GIS Copilot \citep{akinboyewa2024gis} enhances its integration with QGIS, enabling access to QGIS native tools, interface, and external libraries (e.g., GDAL, GeoPandas) through PyQGIS for complex geospatial analysis. 
It incorporates automatic error detection and correction for the coding generation and improves execution reliability, significantly enhancing the usability of existing GIS software.

MapGPT~\citep{zhang2024mapgpt} and GeoGPT~\citep{zhang2024geogpt} leveraged existing frameworks of Langchain to access external map layout and spatial analysis tools. 
The former focuses on map visualization, while the latter focuses on geoanalytical tasks as in this work.
These approaches use the Thought-Action-Observation (TAO) framework, ReAct \citep{yao2023react}, in their implementations. 
However, it is noticed that as the number of tools a single agent must call increases, the error rate rises significantly, complicating the successful execution of tasks. 
Therefore, the use of the TAO framework may lead to suboptimal performance in complex, real-world scenarios~\citep{wang2024tdag}. 
Furthermore, the sequential nature of the ReAct paradigm results in longer task execution times. 
To mitigate these challenges, in this work, we designed a multi-agent framework that can enhance task efficiency and accuracy by assigning each LLM agent only to a specific subtask.

Moreover, almost all studies share the consensus that these tools can enhance productivity and user experience for non-expert users. However, since most of them are reported in the form of demos and example cases in research papers, they lack the opportunity to gather real user feedback. 
Our work differs from these studies in that we provide an initial system that is available for users with the aim of evaluating and addressing user experience through actual user feedback. This makes our research not only theoretically valuable but also more aligned with real-world applications.

\section{Methodology}

The proposed framework, as shown in Figure \ref{fig:framework}, adopts a multi-agent solution. Each agent is responsible for different aspects of semantic analysis and task routing. Each agent can execute complex search tasks by calling various functions without relying on query code generation for geographic calculations.
The entire system was implemented based on OpenAI GPT-4o with the temperature set to zero, where the temperature is a parameter, zero indicating no randomness, resulting in deterministic and consistent responses.

The framework consists of three main components: (1) \textit{Analyzer}, (2) \textit{Explainer}, and (3) \textit{Visualizer}. 
Upon receiving a user prompt, an LLM agent - \textit{Router} - first determines the nature of the task, 
deciding whether the response should involve the visualization of geospatial data or be presented in textual or graphical form.
For example, queries involving geographical entities are categorized as geospatial data-related tasks and routed to the \textit{Analyzer} (see Figure \ref{fig:framework}). 
In contrast, user questions about analysis results or general information on accessible data are routed to the \textit{Explainer}. 
This step has been achieved using a predefined prompt, detailed in Appendix \ref{sec:prompt_router}.


\subsection{Analyzer}
First, an LLM agent - \textit{Relation Analyzer} - performs semantic parsing on the input prompt.
With the entity names and relationships extracted from this agent, the resultant JSON is provided to the \textit{Mission Planner}, another LLM agent, to generate a plan of a sequence of subtasks. 
These steps have been achieved using predefined prompts, detailed in Appendix \ref{sec:prompt_rel_analyzer}.
For instance, for a user query: \textit{"Buildings within 100 meters of the parks in Munich Maxvorstadt"}, the resultant JSON dictionary of \textit{Relation Analyzer} would be structured as follows:

\begin{python}
{
  "entities": [
    {
      "entity_text": "buildings"
    },
    {
      "entity_text": "parks"
    }
  ],
  "spatial_relations": [
    {
      "type": "within 100 meters of",
      "subject": 0,
      "object": 1,
    }
  ],
  "region": "Munich Maxvorstadt"
}
\end{python}

The agent could extract three types of information as dictionary keys, including: \texttt{entities}, \texttt{spatial\_relations}, and \texttt{region}.
The \texttt{entities} key refers to the information about all geographical entities involved in the query. For instance, in this case, the entities include \textit{buildings} and \textit{parks}. 
The \texttt{spatial\_relations} key describes the geographical relationships between entities. Each relation is composed of three keys: \texttt{type}, which defines the type of geographical relation (e.g., "within 100m of"), and \texttt{subject} and \texttt{object}, which indicates the entities involved in the relationship. 
For example, in this query, the spatial relationship specifies that the \textit{buildings} as \texttt{subject} are within 100m of the \textit{park} as \texttt{object}. 
The \texttt{region} key indicates the geographical area constraint for the query, which in this case is limited to the region \textit{Munich Maxvorstadt}.

To decompose the given task into a plan consisting of sequential subtasks, we need to ensure that the agent understands which functions are available.
We have summarized three basic modules for interacting with geospatial data. 
By combining these modules, most of the entry-level geospatial analysis tasks can be addressed. Each module can be called multiple times for different geographic entities to achieve complex retrieval goals. 
In general, the system provides the following types of modules:

\begin{itemize}
\item \textit{Region Selector}: Selects a spatial extent with a bounding box, retrieved based on the extracted region name.
\item \textit{Entity Retriever}: Retrieves geo-entities or groups of geo-entities that match by name across all tables, column names, and database entries.
\item \textit{Data Analyzer}: Filters entities based on specified conditions, including spatial relationships between them.
\end{itemize}

\noindent The descriptions and examples of these three modules are explained to LLM agent - \textit{Mission Planner} - as prompt text (detailed in Appendix \ref{sec:prompt_functions}). 
For the previous prompt example, a task list is generated as follows:

\begin{python}
{
    "Set the bounding box to Munich Maxvorstadt",
    "Get the id_list of parks",
    "Get the id_list of buildings",
    "Filter buildings within 100 meters of the parks",
    "Get the filtered buildings id_list"
}
\end{python}

\noindent The following subsections provide a detailed explanation of each of the three modules in the framework:

\subsubsection{Region Selector}

\begin{figure*}[h]
\centering
\includegraphics[width=.85\textwidth]{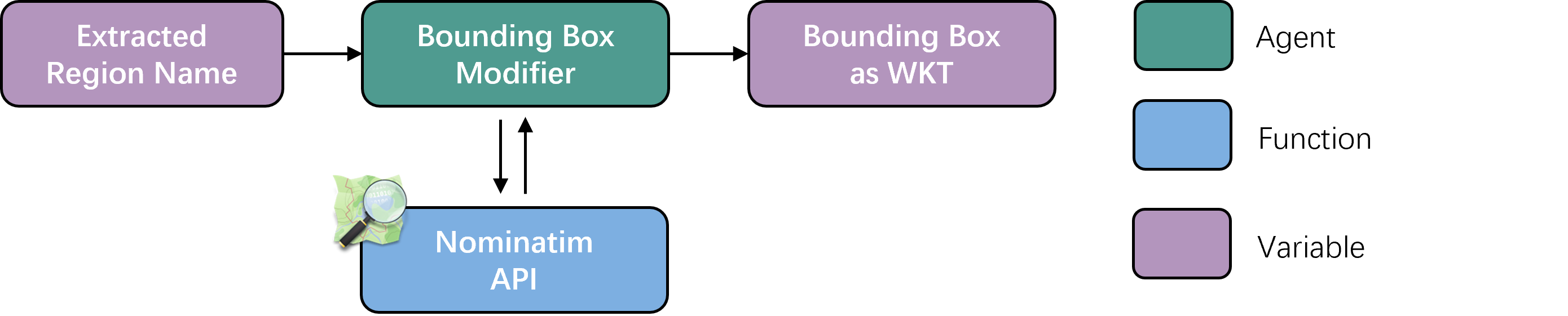}
\caption{Workflow of the module \textit{Region Selector}. }
\label{fig:region_selector}
\end{figure*}

\textit{Region Selector} is one of the fundamental functions in GIS systems, allowing users to define a spatial extent (i.e., a bounding box) to focus on. 
The process is illustrated in Figure \ref{fig:region_selector}.
In this work, users can specify their region of interest by name, aligning queries with specific place names, such as cities or administrative districts. 

To facilitate this, we integrate an existing geocoding API, Nominatim\footnote{Nominatim. \url{https://nominatim.org/}}, which retrieves the corresponding spatial extent based on the user-provided place name. 
This eliminates the need for users to manually look up and upload city boundaries, making the system more suitable for geoportal applications.

After the \textit{Relation Analyzer} extracts the \texttt{region} key from the query, the \textit{Bounding Box Modifier} agent determines the text that should be sent to the geocoding API and evaluates whether modifications are needed to refine the retrieved bounding box according to the user’s intent.
For example, in the case of the previously mentioned query, the identified \texttt{region} key, \textit{Munich Maxvorstadt}, is used to retrieve the bounding box through the API, as no modification is required:
\begin{python}
{
    "bounding_box": 
        [48.139603, 48.157637, 11.538923, 11.588192]
}
\end{python}

In addition, the system allows flexible adjustments to the bounding box based on user prompts. 
Users can expand or shrink the bounding box or select specific sections, such as the eastern or central part. This allows approximate spatial selection without requiring precise coordinates, simplifying the process for non-expert users.
In this case, the \textit{Bounding Box Modifier} agent extracts the place name to be queried by the API and determines whether additional modification is required. 
For example, if \textit{"south of Maxvorstadt"} appears in the \textit{"region"} key, the agent extracts \textit{"Maxvorstadt"} as the place name with:
\begin{python}
{
    "bounding_box": 
        [48.139603, 48.157637, 11.538923, 11.588192]
}
\end{python}
\noindent and then applies a cut to the retrieved bounding box, retaining only its southern part as:
\begin{python}
{
    "bounding_box": 
        [48.139603, 48.148620, 11.538923, 11.588192]
}
\end{python}
This design dynamically adjusts the geographical extent based on user needs. 
The modified extent is stored in WKT format for consistent spatial analysis and reuse within the system.

\subsubsection{\textit{Entity Retriever}}
\label{sec:entity_retriever}

In the context of searching for geographical entities, if the data scope—such as specific tables or columns—is not explicitly mentioned in the query, the retrieval process can generally be abstracted as querying data with two types of text attributes: name and category. 

A common intuitive approach to process such texts with multiple attributes is to jointly encode these attributes, as reported in \cite{chen2019bertjointintentclassification}. The integration of external knowledge enhances the representation of short texts, yielding improved classification performance. However, when applied to complex geographical entities, this approach reveals significant limitations. Many entity names and categories exhibit weak semantic connections, leading to diminished search precision and recall. For example, a geographical name may not inherently indicate its category, and vice versa, resulting in misaligned or incomplete retrieval outcomes. 
These shortcomings are further evidenced in subsequent experiments.

To address the deficiencies of this knowledge-augmented method, a hybrid retrieval approach—combining keyword search and vector search—has been adopted. While this method seeks to enhance the accuracy and recall of text information retrieval, it also presents its own challenges:

\begin{itemize}
    \item \textbf{Interpreting Query Intent Beyond Keywords} \\
    Keywords in a query may not precisely capture the user’s true intent, leading to results that misalign with their expectations. A strict keyword-based approach can overlook contextual nuances, leading to irrelevant matches or missing important results. 

    \item \textbf{Balancing Name and Category Influence} \\
    Vector search processes both name and category attributes simultaneously, but their results may not overlap, making simple intersections or unions ineffective. Name-focused queries, like “schools whose name is related to Ludwig,” prioritize name relevance, while category-focused queries, like “greenery places” emphasize categorical alignment. Ignoring this distinction can lead to missing relevant results or retrieving unrelated ones. To improve accuracy, search strategies must dynamically adjust the weighting of name and category based on the query intent rather than rigidly combining results.


    \item \textbf{Handling Misspellings and Minor Variations} \\
    Exact matches often fail due to misspellings, singular/plural differences, or inconsistent spacing, necessitating the use of vector search. While this approach improves recall, it can also introduce noise and reduce precision, making it crucial to balance flexibility with accuracy in retrieval.  

\end{itemize}

To overcome these persistent issues, a more effective solution integrates the semantic reasoning capabilities of LLM agents with the scalability of vector search. The \textit{Entity Retriever} module is designed to precisely extract geographical entities from user queries within a spatial database, addressing both specific named entities and broader entity categories. This module must accurately interpret the user’s intention—whether it refers to a database table, a column, a category within a column, or a specific entry—despite challenges such as spelling errors, typos, or varying expressions of the same concept (e.g., "water body" encompassing rivers, lakes, or reservoirs). The module supports efficient, flexible, and accurate fuzzy querying by combining vector databases with materialized knowledge graphs. Successful implementation hinges on thorough data preparation. 
In the following subsections, we examine the module from two perspectives: data preparation and workflow.

\begin{figure}[h]
    \centering
    \includegraphics[width=0.48\textwidth, trim={0 150 0 0}, clip]{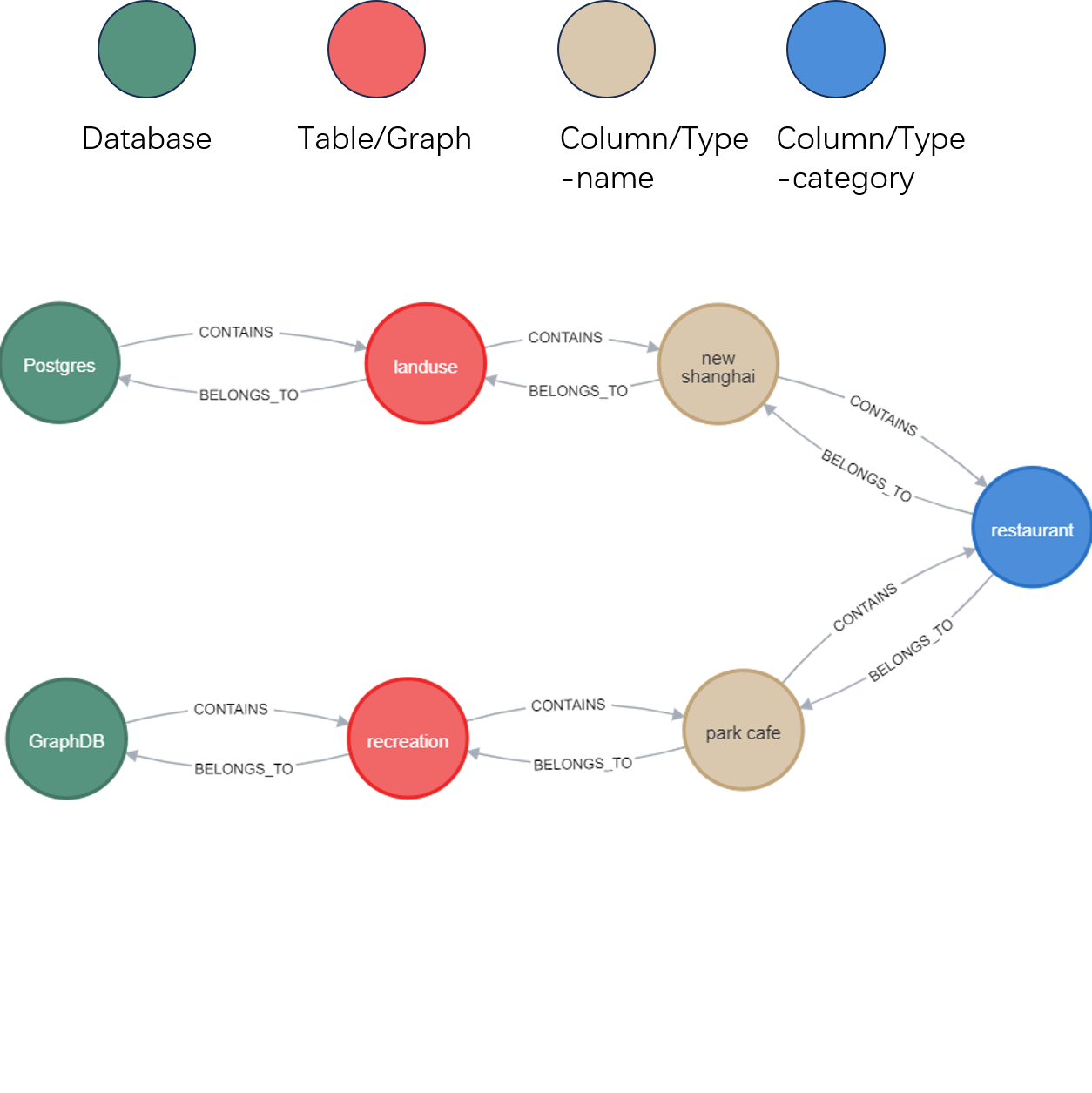} 
    \caption{Relationships between databases in a graph structure. The green nodes indicate database types, while the red nodes correspond to table or graph names. Brown nodes represent column or type names categorized as \texttt{name}, and blue nodes represent column or type names categorized as \texttt{category}. Through these connections, the matched text can be mapped to the corresponding entity information within the database.}
    \label{relationship_graph}  
\end{figure}

\paragraph{\textbf{Data Preparation}}
\label{data_preparation}
During data ingestion, the input data must be analyzed to prepare vector databases and materialized knowledge graphs. 
Vector databases enable semantic search by capturing the meaning and context of unstructured data, whereas the knowledge graph helps pinpoint the location of relevant data by identifying which database it resides in, which table it belongs to, and what the column values are. It also provides the \textit{Explainer} agent in Section \ref{sec:explainer} with structural knowledge of the database for answering general questions about deployed databases. An example of the relationship graph is shown in Figure \ref{relationship_graph}.

Textual column values are first encoded into high-dimensional vectors using the OpenAI text-embedding-3-small model\footnote{OpenAI - Vector embeddings. \url{https://platform.openai.com/docs/guides/embeddings}}.
These vectors are then stored in the Chroma vector database\footnote{Chroma. \url{https://www.trychroma.com/}}. 

We use graphs to show relationships between databases because graph databases are better at handling connections across multiple databases than relational ones. They can efficiently query the storage locations of the same data across different databases through multi-hop searches. Compared to the complex JOIN operations in SQL, graph databases naturally support multi-hop queries, directly connecting concepts, tables, and databases with a MAPS\_TO relationship, resulting in higher query efficiency, which is evidenced in \cite{neo4j_graph_database_speed}. 
Furthermore, graphs serve as an ideal complementary knowledge base for LLMs, as demonstrated in \cite{chen2024llmbasedmultihopquestionanswering} and \cite{chakraborty2024multihopquestionansweringknowledge}. Their natural ability to efficiently handle multi-hop queries not only enhances their integration with LLMs but also simplifies the process of performing complex queries.


\begin{figure*}[h]
\centering
\includegraphics[width=.85\textwidth]{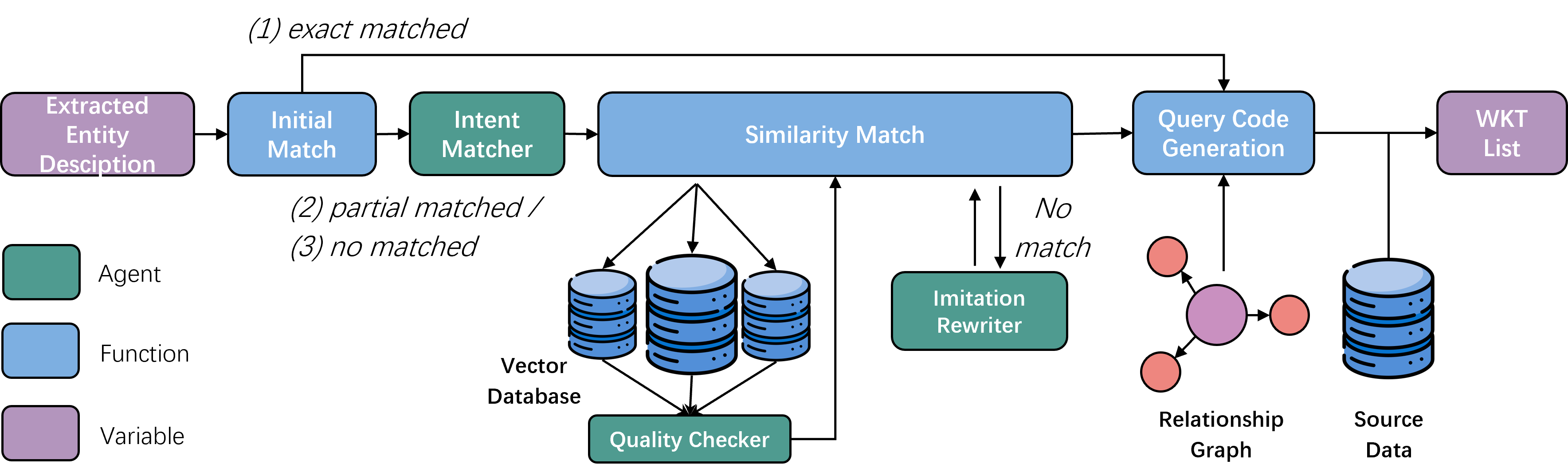}
\caption{Workflow of the module \textit{Entity Retriever}.}
\label{fig:flow_get_entity}
\end{figure*}

\paragraph{\textbf{Module Workflow}}

The workflow for the \textit{Entity Retriever} module is illustrated in Figure \ref{fig:flow_get_entity} and is structured to address three primary query types:

\begin{itemize}
    \item \textbf{Type 1: Data Categories Queries}: Requests related to data categories, such as buildings or parks, requiring matches to database schemas like table names or category names.
    \item \textbf{Type 2: Data Entries Queries}: Queries targeting specific entry names rather than schema elements, such as retrieving records where an entry matches a particular name, necessitating a global search across datasets.
    \item \textbf{Type 3: Knowledge-based Queries}: Requests involving inferred information not explicitly in the database, requiring LLM reasoning. For example, when asked which areas are suitable for farming, the system infers results using soil text descriptions to synthesize insights.
\end{itemize}

\noindent Therefore, the following workflow was developed to handle these scenarios, ensuring that each type of query is appropriately addressed within the system.

When processing user queries, the system first passes the previously extracted entity names to the \textbf{\textit{Initial Match}} function. 
In this function, the system checks whether any keywords (including all table names, category names, and entry names) appear exactly as the input.
All matched terms are standardized to the singular form, ensuring consistency with database naming conventions.  
There are three matching scenarios:
\begin{itemize}
    \item \textbf{Case 1: Exact match}: the entity matches the keyword exactly, such as "building" (table name), "Frauenkirche" (entry name);
    \item \textbf{Case 2: Partial match}: the entity partially matched the keyword, such as "best soil for farming" (matched to table name "soil") or "park named Ludwigstraße" (matched to category name "park");
    \item \textbf{Case 3: No match}: the entity does not match with any keywords, such as "greenery spaces", or "good for planting potatoes".
\end{itemize}

For Case 1 (Exact Match), the matched entity name is passed to the \textit{Query Code Generation} function for direct geospatial data access. 
For both Case 2 and 3, the entity name undergoes a further matching procedure with the vector database. 
The query is passed to the \textbf{\textit{Intent Matcher}} agent, which performs the following tasks:  

\begin{itemize}
    \item \textbf{Table Inference:} The \textbf{\textit{Intent Matcher}} agent first determines whether the queried entity is implicitly related to an existing table. If no relevant table is found, an empty result is also allowed. For example, the query "good for planting potatoes" would be matched to the table "soil".  
    \item \textbf{Category vs. Name Determination:} The agent analyzes the user’s query intent to determine whether it is a Name-focused Search or a Category-focused search. 
    For example, \textit{Isar River} refers to a specific river named \textit{Isar}, requiring a filtered search among all rivers by name. In contrast, \textit{art-related places} represent a category of locations, including \textit{art center}, \textit{museums}, and also places with \textit{art} in their names, such as a café named \textit{Kunst Café} (English: Art Café).  
    \item \textbf{Validation of Partial Matching Results:} The agent evaluates whether the previously matched partial results align with the query intent. If they do, the results are retained. For example, if the matched keyword "soil" corresponds to a table name and is relevant to the query intent, subsequent searches will be constrained within this table's scope.  
\end{itemize}

\noindent Then, using the vector database prepared during the \textit{Preparation} stage, the \textbf{\textit{Similarity Match}} function applies cosine similarity on semantic vectors to compare the queried entity with database entries. 
It identifies the top 50 most relevant entity names or category names, including both entry names (e.g., "Theresienwiese", "Theresienstraße", "Theresienweg" after searching for "Theresien") and category names (e.g., "parks", "forests" after searching for "greenery spaces").

The matched results are then passed to the \textbf{\textit{Quality Checker}} agent for verification and filtering to ensure relevance. 
Empirically, the top 50 matches could cover a sufficient range of categories and entries, which is also the maximum input size allowed by the agent.
Through this process, both Type 1 and Type 2 queries can be successfully addressed.




If no valid results are found in the search, the \textbf{\textit{Imitation Rewriter}} agent reinterprets the user query using the LLM reasoning capabilities to generate new search conditions, performing a search based on the paraphrased input. 
For example, when the query is "Where are the areas with the best soil for farming?" 
(see individual steps of matches in Table \ref{tab:entity_retriever})
, the search is initially restricted to the dataset table named \textit{soil} due to the presence of the word as a table name. 
However, when no exact content related to \textit{the best soil for farming} is found, because the soil table contains only general composition data rather than direct descriptions of farming suitability - the agent reformulates the query based on samples from \textit{soil} table to suggest related information as \textit{Regions with loam soils characterized by rich nutrients, good drainage, and moisture retention}, ensuring a meaningful result. 
The modified prompt is then reprocessed using vector-based calculations only within the \textit{soil} table, retrieving semantically similar records.
This mechanism effectively handles Type 3 queries, involving information not explicitly stored in the database or cases where the similarity threshold is insufficient.
Additionally, caching is employed to avoid redundant processing and maintain system performance.

\begin{table*}[width=2.\linewidth,cols=3,pos=h]
\caption{Examples of intermediate outputs of the \textit{Entity Retriever} module for different match cases.}\label{tab:entity_retriever}
\begin{tabular*}{\tblwidth}{@{} p{4cm} p{6.5cm} p{6cm} @{} }
\toprule

Steps & \textbf{Example 1} & \textbf{Example 2}\\
 & a knowledge-based query matched with & a data query without any match with  \\
 &  a table &  table names or category names\\
\midrule

Extracted entity from Prompt & "areas with the best soil for farming" & "greenery spaces"\\

- 1. Schema Match & Matched candidates: [table:soil, table:area] & Matched candidates: []\\

- 2. Intent Match & Name-focused Search, Valid matches: [table:soil] & Category-focused Search \\

- 3. Similarity Match  & None & Matched categories: [grass, meadow,\\
 &  & greengrocer ...]\\

- 4. Quality Check & - & Valid categories: [grass, meadow ...]\\

- 5. Imitation Rewrite & "Regions with loam soils characterized by rich nutrients, good drainage, and moisture retention" & -\\


\bottomrule
\end{tabular*}
\end{table*}

\begin{figure*}[h]
    \centering
    \includegraphics[width=.85\textwidth]{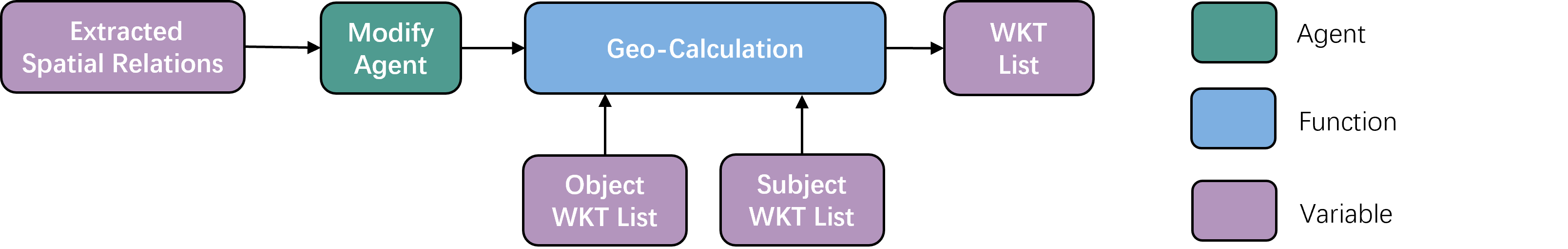}
    \caption{Workflow of the module \textit{Data Analyzer}.}
\label{fig:Geo filter}
\end{figure*}

Lastly, the \textit{Query Code Generation} function then retrieves the schema information of entities from the relationship graph created during the \textit{Preparation} stage and constructs the query code needed to access the source data. 
In the relationship graph, non-spatial data is materialized, enabling the system to precisely locate the database positions of matched entities, including table and column information. 
This allows the system to adapt to various original data storage methods for data access. 
For instance, when Shapefile data is stored in PostgreSQL, the relationship graph helps to transform the matched entities into structured queries (e.g., in this case using SQL), ensuring accurate and context-aware data retrieval. 
Similarly, for user-uploaded data such as GeoJSON, as long as it is included in the relationship graph, the system can make use of the stored schema information and generate corresponding query code based on its database type.

The query results for parks and buildings are organized as follows. Each key in the JSON elements adheres to the format "DatabaseName\_TypeName\_ID", and the corresponding value is the WKT (Well-Known Text) information:

\begin{python}
{
    'land_park_Salinenhof_17978461': <POLYGON (
    (11.579 48.15, 11.579 48.15, 11.579 48...>, 
    'land_park_Maximiliansplatz_144135886': <POLYGON (
    (11.569 48.141, 11.569 48.141, 11.56...>, 
    ...
}
\end{python}
\begin{python}
{
    'buildings_building_Physiotherapie Kinder und 
    Erwachsene_93216444': <POLYGON ((11.577 48.141, 
    11.577 48.141, 11.57...>, 
    'buildings_building_Krone-Villa_153292452': 
    <POLYGON ((11.55 48.145, 11.55 48.145, 11.551 ...>, 
    ...
}
\end{python}

\subsubsection{\textit{Data Analyzer}}


The \textit{Data Analyzer} module performs basic spatial analysis between two sets of geo-entities, using predefined operations including buffer, intersect, contain, and within. The process is illustrated in Figure \ref{fig:Geo filter}
It relies on outputs from the \textit{Relation Analyzer}, which provides information about the subject and object entities, the type of spatial relationship (spatial\_type), whether it involves negation (negation), and any required numerical parameters (e.g., buffer distance).

For example, given the prompt "buildings around 100 meters of a forest," the \textit{Relation Analyzer} identifies "around 100 meters" as spatial relationship  with "buildings" as the subject and "forest" as the object. 
The \textit{Modify Agent} then categorizes this relation as a predefined spatial operation "buffer", with a parameter value of 100, and no negation detected:

\begin{python}
{
    "spatial_type":"buffer",
    "num":100,
    "negation":False
}
\end{python}

\noindent Similarly, for the prompt "buildings outside 100 meters of a forest," the \textit{Relation Analyzer} extracts the spatial relationship "outside 100 meters of" with same subject and object. The \textit{Modify Agent} classifies this as a negation-based operation (inverting the buffer zone) as follows:

\begin{python}
{
    "spatial_type":"buffer",
    "num":100,
    "negation":True
}
\end{python}

\noindent This negation parameter would adjust the computation to include only buildings outside the specified range.

The resultant JSON is then sent to the \textit{Geo-Calculation} function.
All spatial analysis operations are pre-implemented using Shapely’s geometric operations\footnote{\url{https://shapely.readthedocs.io/}} and include relevant descriptions so that the agent can call directly. The function’s efficiency is enhanced by using Shapely’s STRtree (Sort-Tile-Recursive tree) spatial index. STRtree, based on R-tree principles, organizes geometry bounding boxes into a hierarchical tree structure, enabling fast queries for geometries that intersect specific areas or objects. By pre-implementing additional geometric operations, further types of spatial analyses can be easily incorporated into this framework.

This module processes geo-entities in WKT list format as both input and output. This design allows results to be directly reused in subsequent \textit{Data Analyzer} operations as subject or object input, enabling iterative refinement by adding constraints through follow-up queries.
Additionally, this consistency enhances performance efficiency, particularly by reducing the token usage of the LLM model.

\subsection{Explainer}
\label{sec:explainer}
\begin{figure*}[h]
    \centering
    \includegraphics[width=0.85\textwidth]{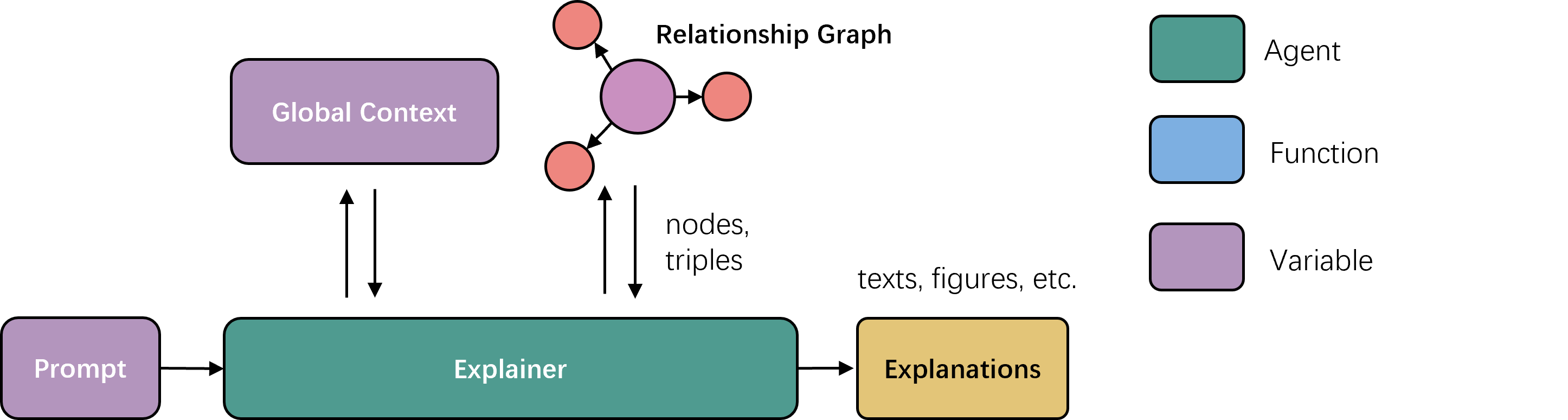}
    \caption{Workflow of the \textit{Explainer}}
    \label{fig:explainer_structure}
\end{figure*}

If the \textit{Router} identifies the user's prompt as a general explanatory question, as demonstrated in Appendix \ref{sec:prompt_router}, it sends the prompt to the \textit{Explainer} for further interpretation. 
The structure of the \textit{Explainer} is illustrated in Figure \ref{fig:explainer_structure}, while its prompt is discussed in Appendix \ref{sec:prompt_explain}.
It is designed to handle two scenarios: 
\begin{itemize}
    \item it interprets the outputs or intermediate results from the \textit{Analyzer} and presents a descriptive answer or visualizes the data using charts and graphs; 
    \item it addresses general inquiries related to the scope of queryable information, such as information of databases. 
\end{itemize}

In the first scenario, since the context of the \textit{Explainer} is a collection of the global context of each agent, it can obtain information about everything happening within the entire framework, including all variables stored by the \textit{Analyzer} during the search. Therefore, it can provide detailed explanations to the user or utilize these variables to generate charts.
For example, after asking "all buildings in Maxvorstadt", then a user inputs “Can you draw me a diagram for area distribution of buildings you searched?”, the \textit{Router} instructs the \textit{Explainer} to find related variable and generates Python code to create a histogram using the WKT list as input.

In the second scenario, the \textit{Explainer} can search for edges and nodes in the relational knowledge graph to retrieve database-relevant information for users. This process is facilitated through Cypher code generation.
For example, if a user asks \textit{"what are the datasets we have?"}, It would attempt to implement code to search for information stored in the relationship graph and then interpret the results for the user.
%
%
Based on the retrieved nodes and edges, the \textit{Explainer} agent could generate an integrated response as follows: \\

\begin{python}
"Please tell me which one or more of these datasets you 
are interested in and what information you want to query.

You have the following datasets in your database:
1. soil
2. roads
3. points
4. area
5. buildings"
\end{python}

\noindent The \textit{Explainer} process may last for several iterations until it determines that the task is complete, at which point it will terminate and provide an explanation. This approach is effective for handling problems that require deep understanding.




\subsection{Visualizer}
The \textit{Visualizer} module supports visualization of the outputs (i.e., a list of WKT objects) of individual subtasks generated by the \textit{Analyzer} module. Each output is linked with a button within the subtask list shown in the chat window. This allows users to view the map visualization of results at any intermediate step. 

The entire module is realized using \textit{leaflet}\footnote{Leaflet - a JavaScript library \url{https://leafletjs.com/}} as shown in Figure \ref{fig:interface}.
Each output list may include data from multiple tables, which are displayed in layers for better categorization. Users can toggle the visibility of individual layers by clicking on their names. To enhance visualization, adjacent geographic entities are distinguished using random colors, and a default level of transparency is applied to avoid obscuring the base map or other data. 
For interaction, hovering the mouse over an element in the visualization displays its name and basic information, providing an intuitive user experience.

\section{Experiments and Results}


In this section, we deploy our system (Sec. \ref{sec:deployment}) and present a series of case studies (Sec. \ref{sec:case_studies}) showcasing what queries the system is able to answer. 
Furthermore, a series of experiments were performed to evaluate the system (Sec. \ref{sec:sys_eval}) as well as the user experience (Sec. \ref{sec:user_tests}).

\subsection{System Deployment}
\label{sec:deployment}

The system deployment will be introduced from the following two aspects: the datasets used in this work and the process of data import.

\subsubsection{Datasets}

The dataset used for this study and evaluation is mainly sourced from OpenStreetMap (OSM), specifically the version provided by Geofabrik for Upper Bavaria\footnote{Download OpenStreetMap data for this region: Oberbayern. Source: \url{https://download.geofabrik.de/europe/germany/bayern/oberbayern.html}}.
The dataset includes multiple shapefiles, each representing different OSM layers such as buildings, land use, road and railway networks, and points of interest. 
Compared to the original PBF-format files, these shapefiles have undergone a data aggregation process into a structured layered GIS format\footnote{OpenStreetMap Data in Layered GIS Format. Source: \url{https://www.geofabrik.de/data/geofabrik-osm-gis-standard-0.7.pdf}}, making them more suitable for our use.

In addition to the general topographical data, domain-specific data could also been utilized in our system. 
This includes soil data from the Bavarian State Office for Environment, Health, and Consumer Protection\footnote{ Bavarian State Ministry of the Environment and Consumer Protection | Bayerisches Staatsministerium für Umwelt und Verbraucherschutz. Source: \url{https://www.stmuv.bayern.de/english/index.htm}}, specifically the Overview Soil Map of Bavaria (1:25,000)\footnote{Übersichtsbodenkarte 1:25.000. Source: \url{https://www.lfu.bayern.de/boden/karten_daten/uebk25/index.htm}}, which is available as open data under the CC BY 4.0 license. 
The soil types are represented using a category code along with a text description in German, as illustrated in Table \ref{tab:soil_example}. 
This dataset contains specialized domain knowledge, making it challenging for non-expert users as well as those unfamiliar with the German language.

\begin{table*}[width=2.\linewidth,cols=2,pos=h]
\caption{Examples of soil categories and descriptions.}\label{tbl:react}
\label{tab:soil_example}
\begin{tabular*}{\tblwidth}{@{} LL@{} }
\toprule
Original class and textual description & English Translation\\
\midrule
\textbf{997b}: Besiedelte Flächen mit anthropogen überprägten & \textbf{997b}: Populated areas with anthropogenically influenced\\
Bodenformen und einem Versiegelungsgrad < 70 \%;  & soil forms and a sealing degree < 70 \%; not differentiated\\
bodenkundlich nicht differenziert &  in terms of soil science \vspace{2mm}\\
\textbf{21}: Fast ausschließlich humusreiche Pararendzina aus  & \textbf{21}: Almost exclusively humus-rich Pararendzina from  \\
Carbonatsandkies bis -schluffkies (Schotter), gering & carbonate sand gravel to silt gravel (gravel), sparsely  \\
verbreitet mit flacher Flussmergeldecke & distributed with flat river marl cover\\
\bottomrule
\end{tabular*}
\end{table*}

\subsubsection{Data Import and System Deployment}

Our framework performs geospatial relationship operations in memory. The design of the relationship graph enables efficient cross-computation between different data locations. As a result, our framework can easily import new data, such as JSON files. When users upload data, text data and file locations are stored in the vector database and the relationship graph, as described in Section \ref{data_preparation}.

\begin{figure*}[h]
    \centering
    \includegraphics[width=0.95\textwidth]{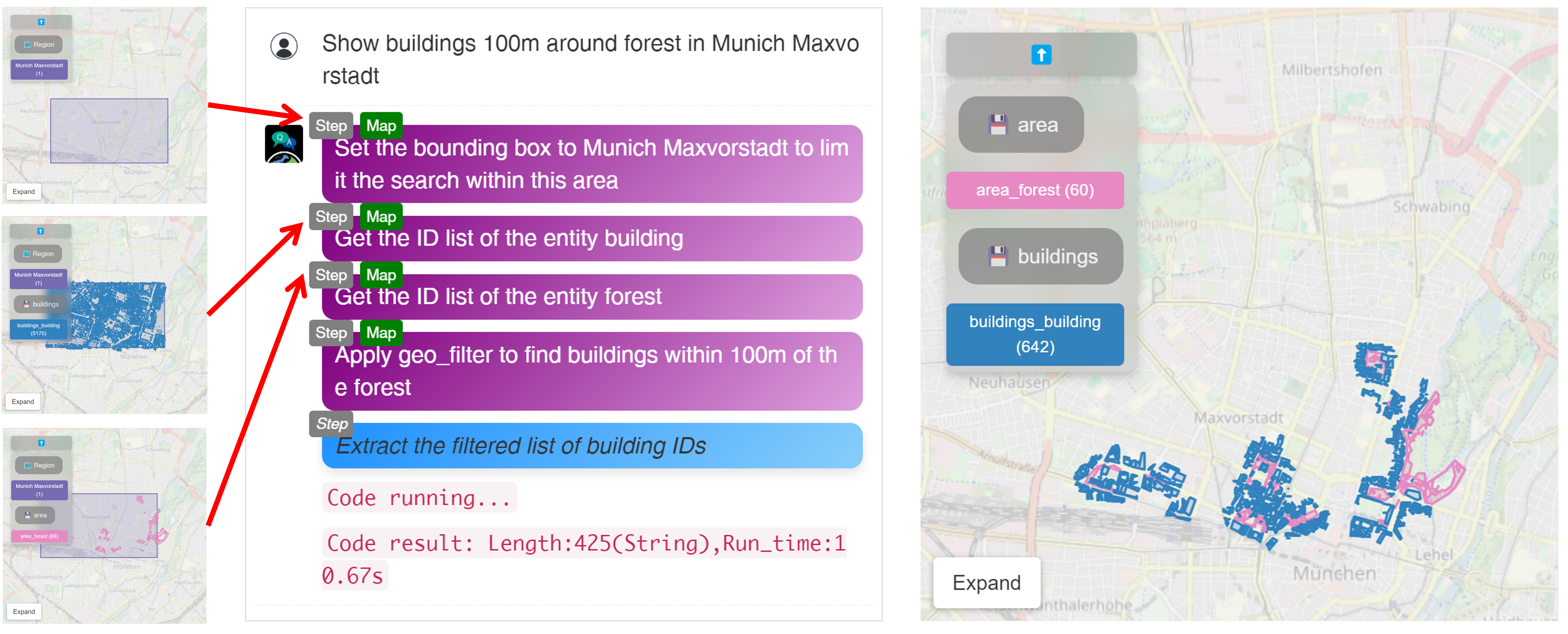}
    \caption{The system user interface, and the step-by-step results of the system when a single map button is clicked.}
\label{fig:interface}
\end{figure*}

\subsection{Case Studies}
\label{sec:case_studies}

Five common scenarios are selected to demonstrate how our proposed system effectively addresses questions of varying complexity:

\subsubsection{Basic Spatial Retrieval Task}
\begin{figure}[h]
    \centering
    \begin{subfigure}[b]{0.49\textwidth}
    \centering
    \includegraphics[width=\textwidth]{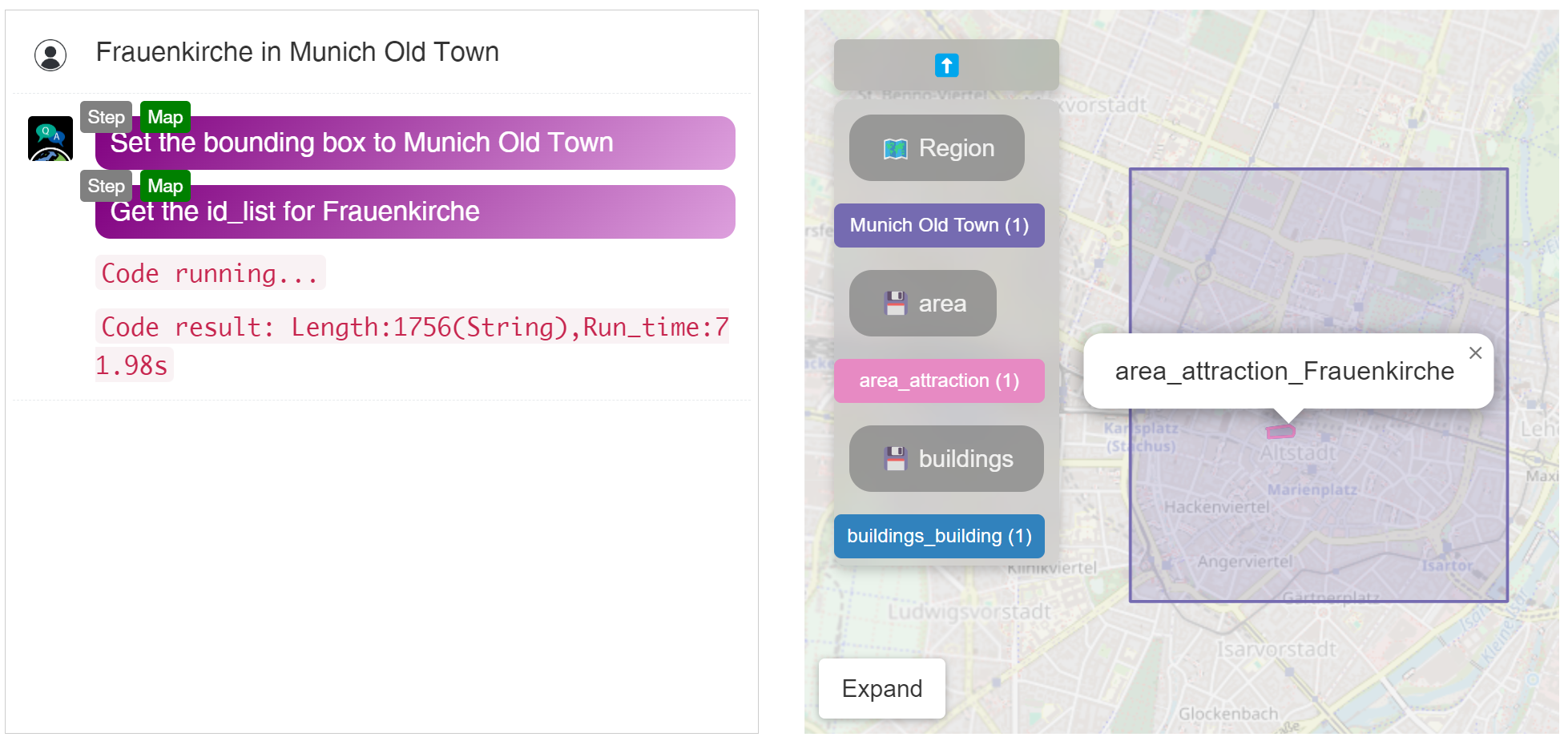}
    \caption{Search for certain objects by name: \textit{Frauenkirche in Munich Old Town}}
    \label{fig:search_items}
    \end{subfigure}
    \hfill
    \begin{subfigure}[b]{0.49\textwidth}
        \centering
        \includegraphics[width=\textwidth]{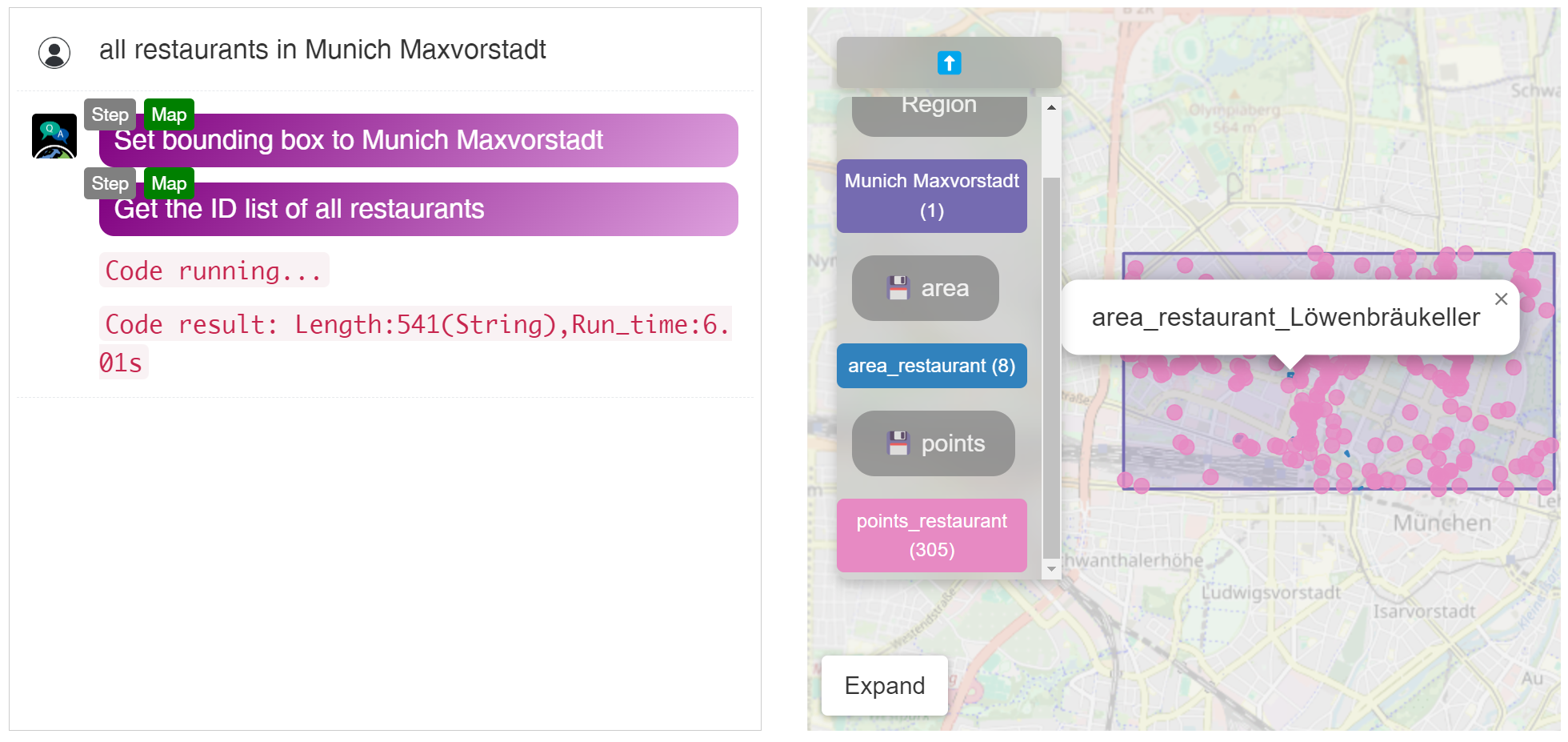}
        \caption{Search for groups of objects: \textit{All restaurants in Munich Maxvorstadt}}
        \label{fig:search_groups}
    \end{subfigure}
    \caption{Example of retrieval question for items by name or for groups.}
    \label{fig:retrieval_question}
\end{figure}

The proposed system is capable of handling basic spatial retrieval queries by allowing users to provide search criteria, enabling them to efficiently identify specific objects or groups of objects within a designated area and visualize their locations. For instance, Figure \ref{fig:search_items} shows a case of retrieving a single object by name, such as finding the “Frauenkirche” in Munich’s Old Town. Two results have been visualized, one is the attraction named \textit{Frauenkirche}, and the other is a building named the same.
Meanwhile Figure \ref{fig:search_groups} illustrates retrieving a group of objects, like “all restaurants in Munich Maxvorstadt.” 

\subsubsection{Spatial Analysis Task}
In addition to simple searches, our system can analyze relationships between multiple spatial objects, such as distance, overlap, or proximity.
Figure \ref{fig:interface} showcases how users can perform spatial analysis through an intuitive interface. 
In this example, we requested the system to identify buildings located within 100 meters of a forest in Munich's Maxvorstadt district, with the results clearly highlighting the buildings and forests. By simply clicking a button on the map, the system provides step-by-step intermediate results, allowing users to efficiently and intuitively verify whether the results are accurate.

\subsubsection{Follow-Up questions for multi-criteria spatial analysis}

\begin{figure}[h]
    \centering
    \begin{subfigure}[b]{0.51\textwidth}
        \centering
        \includegraphics[width=\textwidth]{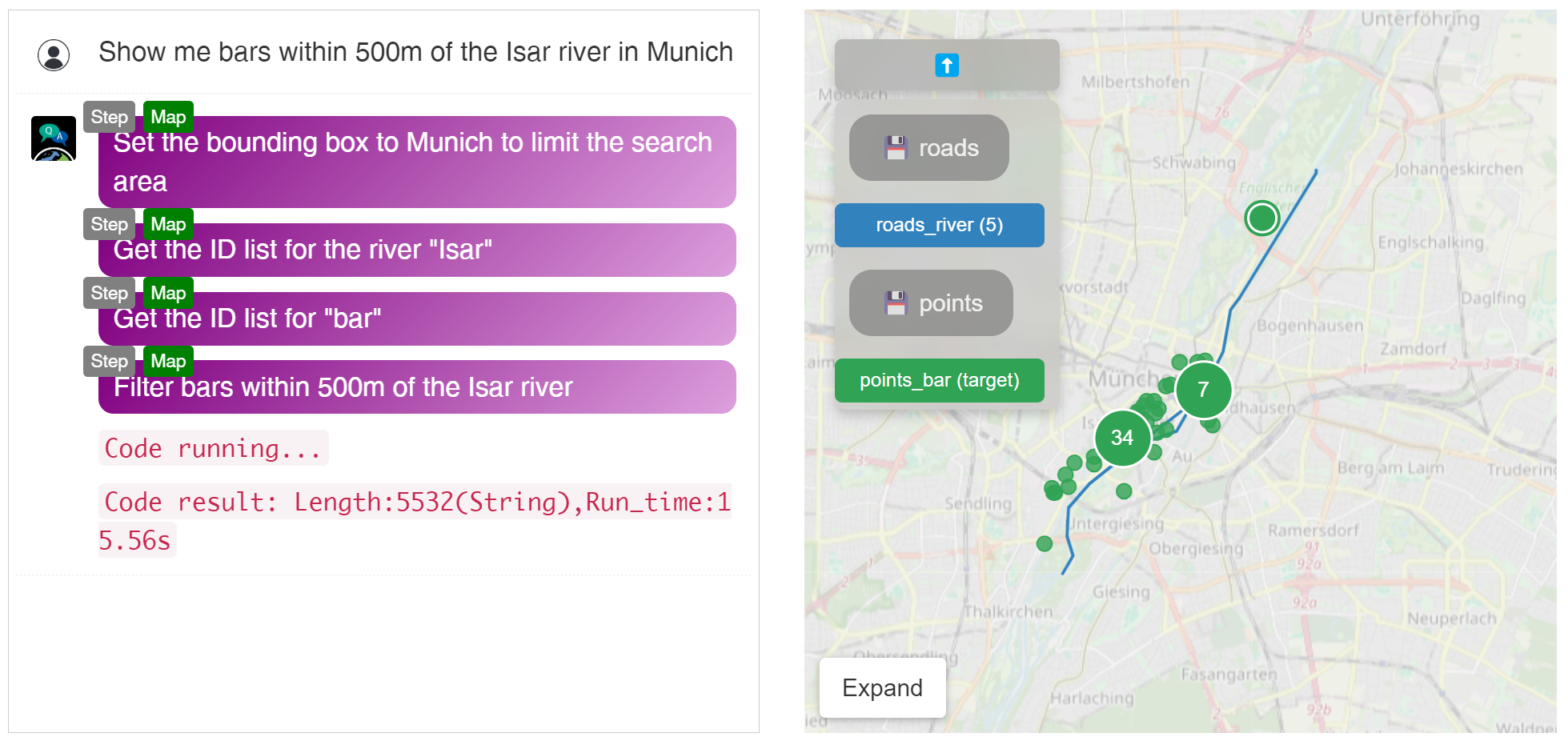}
        \caption{Show me bars within 500m of the Isar river in Munich.}
        \label{fig:first_image}
    \end{subfigure}
    \hfill
    \begin{subfigure}[b]{0.51\textwidth}
        \centering
        \includegraphics[width=\textwidth]{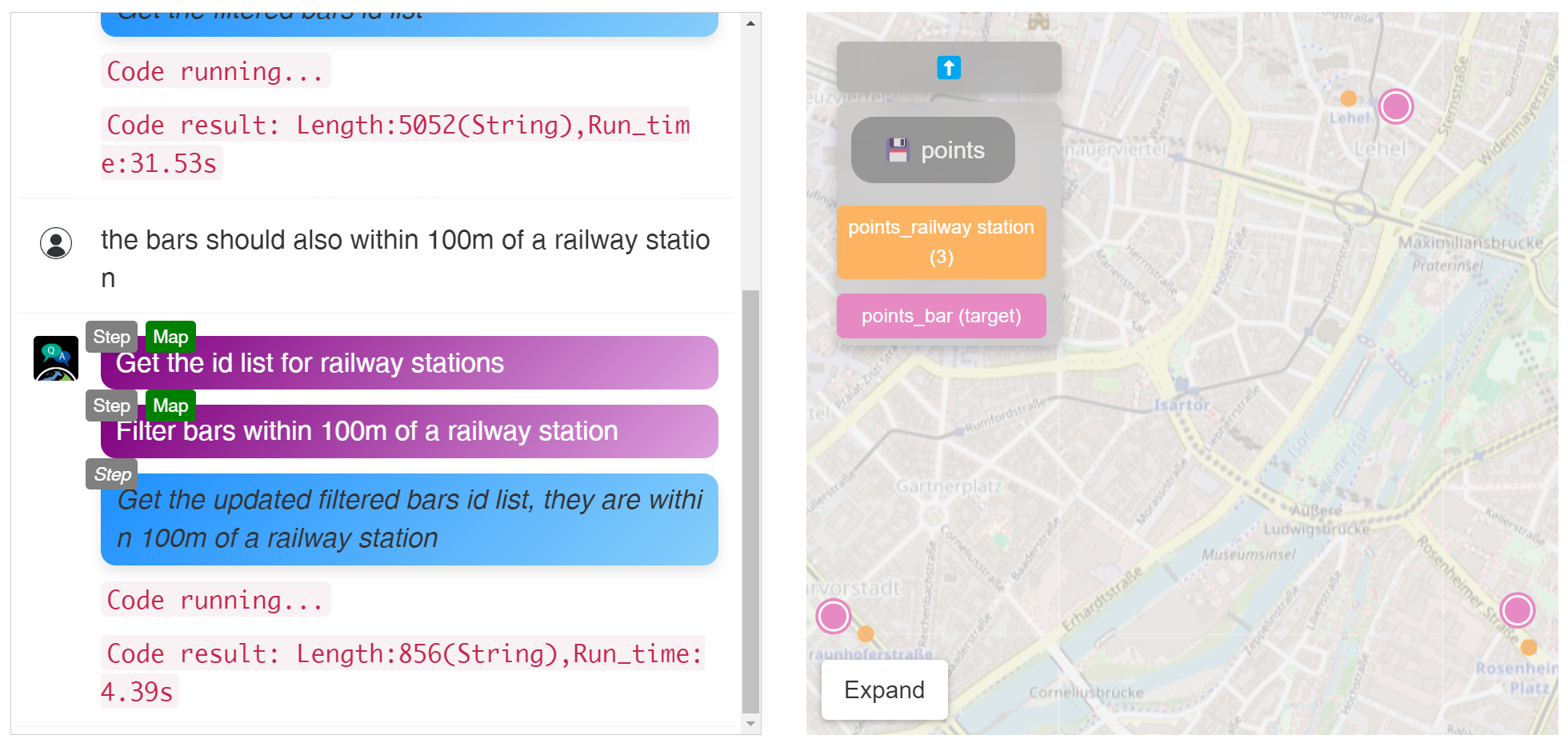}
        \caption{The bars should also within 100m of a railway station.}
        \label{fig:second_image}
    \end{subfigure}
    \caption{Example of a follow-up question for multi-criteria spatial analysis.}
    \label{fig:followup_question}
\end{figure}

In practical scenarios, users often pose follow-up questions to the spatial analysis results that refine their initial queries with additional conditions.
The initial query asks the system to \textit{show bars within 500m of the Isar river in Munich”} (Figure \ref{fig:first_image}), followed by a refined follow-up question specifying that “the bars should also be within 100m of a railway station” (Figure \ref{fig:second_image}). This question can be further refined by users.

\subsubsection{Integrating LLM Knowledge}
\begin{figure}[h]
    \centering
    \begin{subfigure}[b]{0.5\textwidth}
        \centering
        \includegraphics[width=\textwidth]{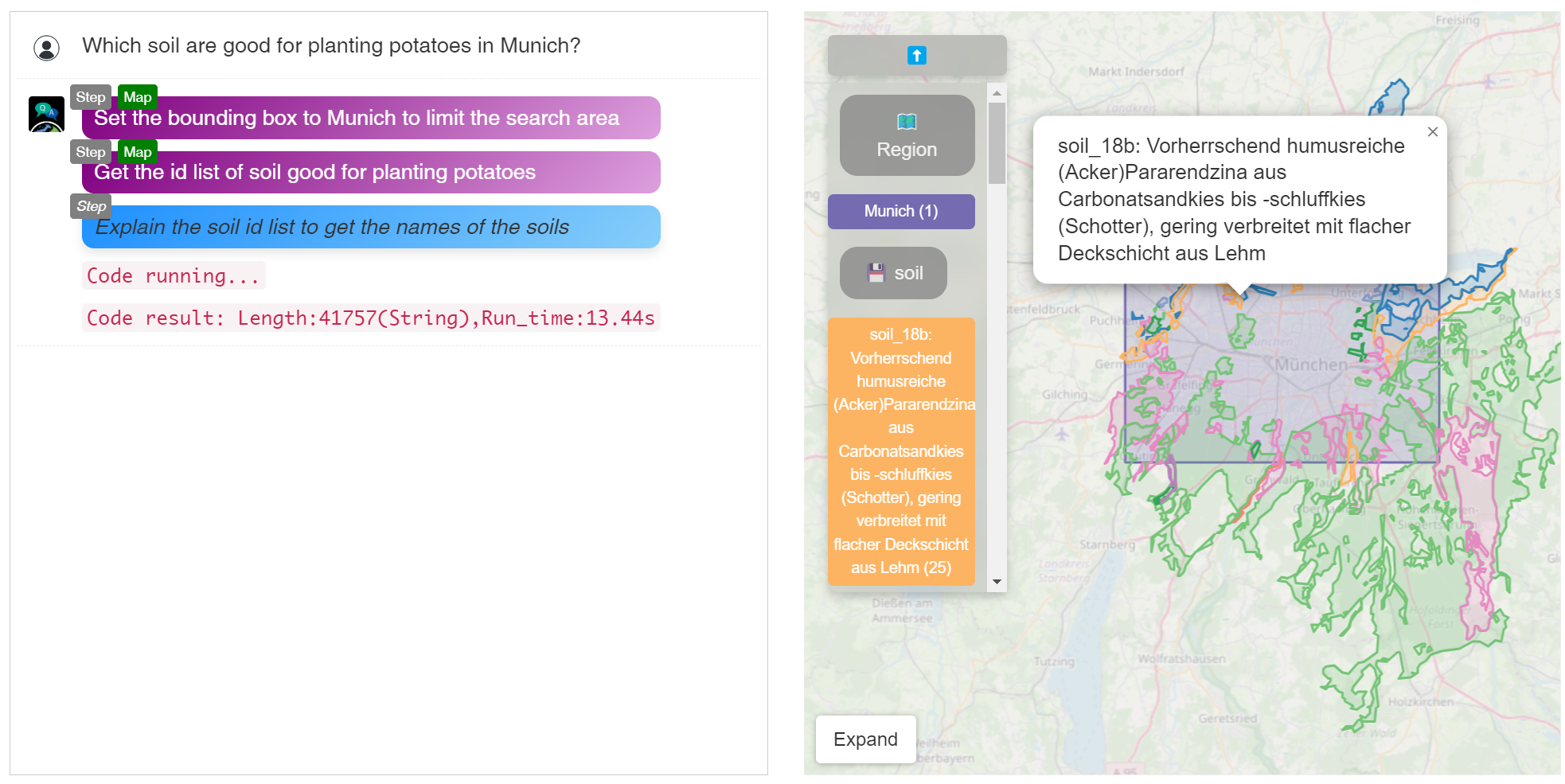}
        \caption{Questions in need of external knowledge: \textit{Which soil is good for planting potatoes in Munich?}}
        \label{fig:soil_ques}
    \end{subfigure}
    \hfill
    \begin{subfigure}[b]{0.5\textwidth}
        \centering
        \includegraphics[width=\textwidth]{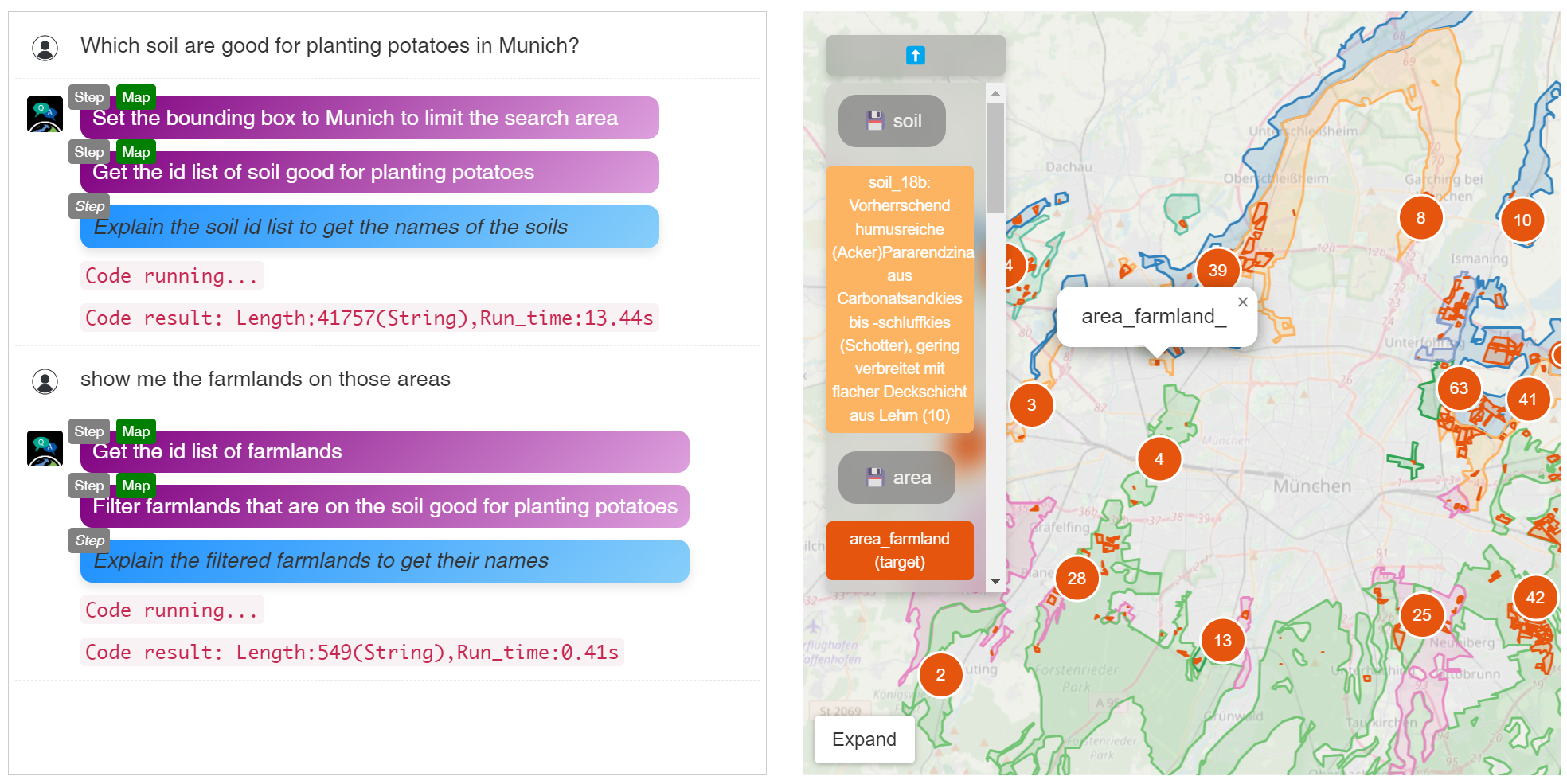}
        \caption{Questions in need of external and internal knowledge: show me the farmlands on those areas}
        \label{fig:farm_soil}
    \end{subfigure}
    \caption{Example of a follow-up question for multiple condition spatial analysis.}
    \label{fig:gpt_know_question}
\end{figure}

In many cases, the information we look for is not explicitly recorded as an entry in the database. For example, if we want to determine suitable locations for growing potatoes based on the system's built-in soil data (Figure \ref{fig:soil_ques}), the database may only contain detailed descriptions of soil types without explicitly stating whether they are suitable for potato. 
By leveraging the knowledge of LLM, the system can interpret the descriptions of soil characteristics and infer the best locations for planting, effectively bridging the gap between raw data and practical decision-making. 
Subsequently, the answer can be combined with the information in our database to identify farms with the corresponding soil types (Figure \ref{fig:farm_soil}).

\subsubsection{Descriptive Questions}
\begin{figure}[h]
    \centering
    \includegraphics[width=0.49\textwidth]{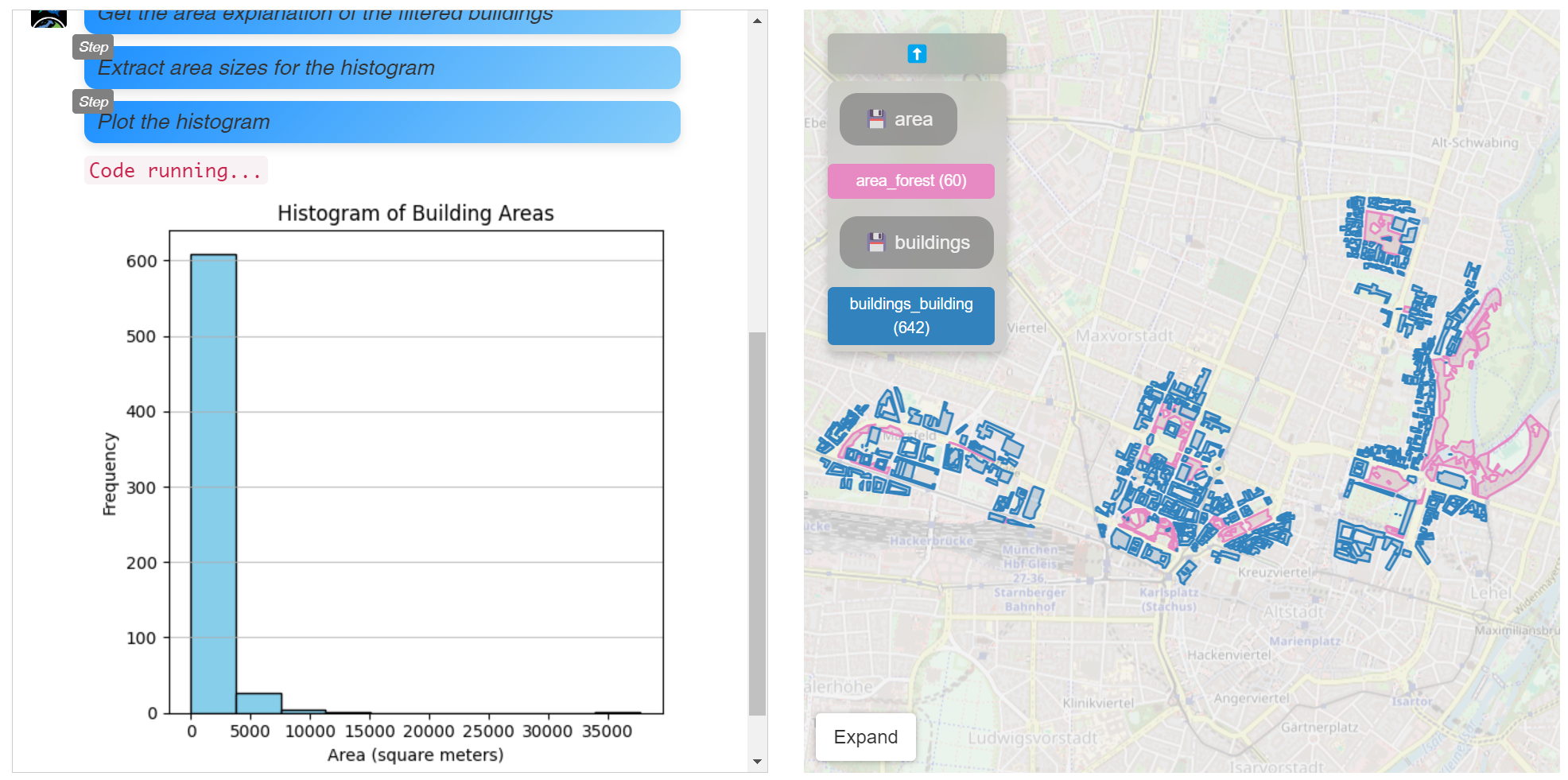}
    \caption{Descriptive Question: \textit{show me a histogram of their area}}
\label{fig:show_histogram}
\end{figure}

Beyond the above questions that require visualization of geographic entity, the system can also leverage the agent’s language understanding and programming capabilities to enhance result descriptions or generate additional figures. This allows for more comprehensive and insightful responses, such as summarizing  or creating a histogram showing the area distribution of the retrieved buildings (Figure \ref{fig:show_histogram}).

\subsection{System Performance Evaluation}
\label{sec:sys_eval}

For a Q\&A system, the most critical evaluation metric is its ability to correctly answer questions. Using LLMs to generate questions specific to the database content allows for an effective evaluation of the system's understanding and retrieval capabilities, ensuring that it accurately extracts and processes relevant information. 
We designed a systematic evaluation process, as shown in Figure~\ref{fig:evaluation}. 

\subsubsection{Dataset Generation}

First, we randomly selected geographic entity types (e.g., parks, rivers, or buildings), specific entity names (e.g., Isar River), and their spatial relationships (i.e., among contains, intersect, within, and buffer), combining them in various ways.
These combinations are then used to generate SQL queries. If a generated SQL query returns a valid result from the database, it is considered a valid query for our dataset.
We then employed an LLM to rephrase these combinations into natural language queries in multiple ways (i.e. the step  of \textit{Rewrite}), ensuring the system could handle diverse expressions and adapt to different phrasings.
Finally, the natural language queries, along with their corresponding SQL queries and results, were added to our ground truth dataset for the follow-up comparison.
In the following, we demonstrate how varying difficulty levels for both query tasks and spatial analysis tasks were evaluated:

\paragraph{Query}
Two difficulty levels were established:

\begin{figure*}[ht]
    \centering
    \includegraphics[width=\linewidth]{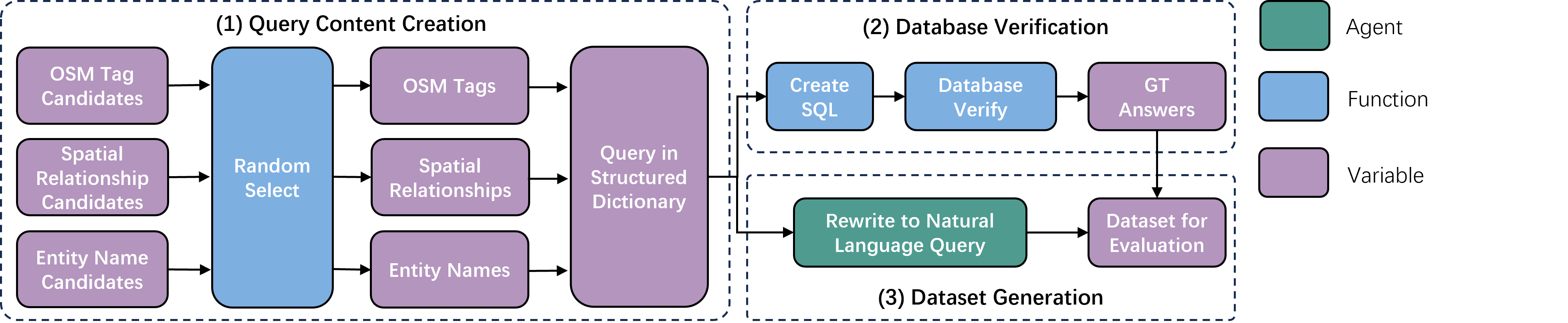}
    \caption{The evaluation process of Q\&A System.}
    \label{fig:evaluation}
\end{figure*}

\begin{enumerate}[label={(\arabic*)}]

\item Exact Match: the original wording from the database appears in the query.

- \textit{"road called Arcisstrasse"}

- \textit{"building named Alte Pinakothek"}

- \textit{"all roads in Munich Moosach"}

- \textit{"all buildings in Munich Maxvorstadt"}

\item Paraphrased Expression: the original wording does not appear, but the query is reworded in a different way. This was achieved by modifying the prompt of the "Rewrite" agent to accommodate this requirement.

- \textit{"Where can I get my vehicle cleaned in the city?"}, where the given category is "car wash".

- \textit{"Where can I find carpooling services around me?"}, where the given category is "car sharing".

\end{enumerate}

\paragraph{Spatial Analysis}

By randomly combining different geographic entities and their spatial relationships, we progressively increased the complexity of queries. 
This allowed us to assess the system's performance across various scenarios and evaluate its ability to accurately interpret and analyze queries involving spatial relationships.
Moreover, spatial analysis problems often involve geographic entities that are constrained by descriptive modifiers. 
Given that the dataset fields vary - with only the "name" field being universally present - we standardized our approach by exclusively utilizing name modifiers for data generation of the following four example scenarios:

\begin{enumerate}[label={(\arabic*)}]

\item Two-Class Spatial Analysis without Specified Names: Each query involves two types of geo-entities and one spatial relationship, without specifying any names.

- Example: \textit{"Can you show me the recreation ground that contains a kindergarten?"}. In this query, the entities types "recreation ground" and "kindergarten" are used without specific modifiers, and the spatial relationship “contains” links them.


\item Two-Class Spatial Analysis with Specified Names: Each query contains two types of geo-entities, both of them specifying the names.  

- Example: \textit{"I want to know if there is a park named Luitpoldpark that contains a pitch named Basketball in Munich"}. In this query, the geo-entity types “park” and “pitch” are used, and each is specified by its name. The spatial relationship “contains” is applied to link the two entities.


\item Three-Class Spatial Analysis with Specified Names: Each query contains three types of geo-entities, all of them specifying the names.  

- Example: \textit{"I'm curious to find clothes shops that are located within 500 meters of the tertiary road known as Westendstraße, as well as within 500 meters of the tertiary road called Barthstraße."}. In this query, three geo-entity types are involved: the clothes shop and two tertiary roads, each specified by name. The spatial constraints—requiring the shop to be within 500 meters (i.e. "buffer") of both roads—assess the system’s ability to simultaneously handle multiple spatial relationships. 


\item Four-Class Spatial Analysis with Specified Names: Each query contains four types of geo-entities, all of them specifying the names. 

- Example: \textit{"do you know where I can find the journey of the river called Rote Traun as it gracefully intersects with streams named Wiener Graben, Sulzbach, and Falkenseebach?"}. In this query, four geo-entities are involved: one river and three streams, each specified by name. The spatial relationship “intersects” is used to describe how the river’s path interacts with each of the streams, thereby testing the system’s ability to manage multiple spatial relationships concurrently.


\end{enumerate}

\subsubsection{Evaluation Results}
The evaluation results from the same aforementioned aspects are presented here.

\paragraph{Query}
In this task, the baseline method used for comparison, as described in Section \ref{sec:entity_retriever}, concatenates the "name" and "category" fields of each entity for vector search, following the approach outlined in \cite{chen2019bertjointintentclassification}. 
The evaluation focuses solely on the retrieval ability of \textit{Entity Retriever}.

As shown in Table \ref{tab:search_performance}, our approach consistently outperforms the baseline in both precision and recall. Precision is defined as the proportion of retrieved entities that are not ground truth entities, while recall measures the proportion of ground truth entities that are not retrieved. 
Since rephrased geo-entities may introduce additional related content, we manually annotate the (2) Paraphrased Expression dataset to ensure the ground truth is accurately defined and contextually appropriate.

By integrating a multi-agent LLM design, our method achieves more precise matching and enhances retrieval comprehensiveness. Compared to the baseline, which relies solely on vector search, our approach effectively mitigates the key challenges of determining an appropriate similarity threshold and the negative impact of semantic inconsistencies between different types of labels, such as names and categories.


\begin{table*}[width=1.8\linewidth,cols=6,pos=h]
\caption{Comparison of Search Performance Between Baseline and Ours.}
\label{tab:search_performance}
\begin{tabular*}{\tblwidth}{@{} Lccccc@{}}
\toprule
\multirow{2}{*}{\textbf{Dataset details}} &  \textbf{\# of}  & \multicolumn{2}{c}{\textbf{Baseline}} & \multicolumn{2}{c}{\textbf{Ours}} \\
 & \textbf{Queries} & Precision & Recall & Precision & Recall \\
\midrule
(1) Exact Match  & 100 &  95\% & 68\% & \textbf{100\%} & \textbf{100\%} \\
(2) Paraphrased Expression & 100 & 84\% & 45\% & \textbf{95\%} & \textbf{78\%} \\
\bottomrule
\end{tabular*}
\end{table*}

\paragraph{Spatial Analysis}

\begin{table*}[width=2\linewidth,cols=7,pos=h]
\caption{Accuracy Comparison of Responses for Tasks of Different Difficulty and Baseline Methods}
\label{tab:correct_assess}
\begin{tabular*}{\tblwidth}{@{} LCCCCC@{} }
\toprule
\textbf{Dataset details} & \textbf{\# of} & \textbf{Baseline 1} & \textbf{Baseline 2} & \textbf{Baseline 3}   & \textbf{Ours}\\
 & \textbf{Queries} & (SQL) & (SQL with template) & (Single Agent with ReAct) & \\
\midrule

(1)  2 geo-entities without modifiers & 100 & 74\% & 76 \% & 41\% & \textbf{96\%} \\
(2)  2 geo-entities with modifiers & 100 & 65\% & 72 \% & 36\% & \textbf{92\%} \\
(3)  3 geo-entities with modifiers & 100 & 70\% & 83 \% & 59\% & \textbf{95\%} \\
(4)  4 geo-entities with modifiers  & 100 & 61\% & 87 \% & 52\% & \textbf{90\%} \\
\bottomrule
\end{tabular*}
\caption{Comparison of Average Token Usage for Tasks of Different Difficulty and Baseline Methods (Input represents input token count, and Output represents Output token count, and the total token count is shown above them.)}
\label{tab:tokens}
\begin{tabular*}{\tblwidth}{@{} Lccc|cc|cc|cc@{} }
\toprule
\multirow{2}{*}{\textbf{Dataset details}} & \textbf{\# of}  & \multicolumn{2}{c}{\textbf{Baseline 1}} & \multicolumn{2}{c}{\textbf{Baseline 2}} & \multicolumn{2}{c}{\textbf{Baseline 3}} & \multicolumn{2}{c}{\textbf{Ours}} \\
& \textbf{Queries} & \multicolumn{2}{c}{(SQL)} & \multicolumn{2}{c}{(SQL with template)} & \multicolumn{2}{c}{(Single Agent with ReAct)} &  \multicolumn{2}{c}{} \\
&  & Input & Output & Input & Output & Input & Output &  Input & Output\\
\midrule

(1) 2 geo-entities & \multirow{2}{*}{100} & \multicolumn{2}{c}{365.61} & \multicolumn{2}{c}{551.48} & \multicolumn{2}{c}{2727.75} &  \multicolumn{2}{c}{430.12} \\
 without modifiers &  & 229.35 & 136.26 & 409.35 & 142.13 & 2399.03 & 328.72 &  345.0 & 85.12 \\

(2) 
2 geo-entities  & \multirow{2}{*}{100} & \multicolumn{2}{c}{310.59} & \multicolumn{2}{c}{477.77} & \multicolumn{2}{c}{2239.99} &  \multicolumn{2}{c}{448.96} \\
with modifiers &  & 208.66 & 101.93 & 342.96 & 134.81 & 1891.75 & 348.24 &  345.0 & 103.96 \\

(3) 3 geo-entities  & \multirow{2}{*}{100} & \multicolumn{2}{c}{378.64} & \multicolumn{2}{c}{550.72} & \multicolumn{2}{c}{4266.95} &  \multicolumn{2}{c}{536.89} \\
with modifiers &  & 228.18 & 150.46 & 376.18 & 174.54 & 3742.77 & 524.18 &  345.0 & 191.89 \\

(4) 4 geo-entities  & \multirow{2}{*}{100} & \multicolumn{2}{c}{478.15} & \multicolumn{2}{c}{902.31} & \multicolumn{2}{c}{6655.70} &  \multicolumn{2}{c}{619.48} \\
with modifiers &  & 258.29 & 219.86 & 617.29 & 285.02 & 5944.22 & 711.48 &  345.0 & 274.48 \\

\bottomrule
\end{tabular*}
\end{table*}

We compared the following baselines:

\begin{itemize}
\item Baseline 1: LLM-Based SQL Generation for Database Queries

The most intuitive method to query a database using natural language is to generate SQL with LLM, which we adopt as our baseline, similar to the approaches summarized in Section \ref{sec:llm_query}. 
Since it's impractical to provide the entire database to the LLM, we simplify the process by providing the tables and column names relevant to every single example query to LLM, making SQL generation feasible.

\item Baseline 2: LLM-Based SQL Generation with Templates for Database Queries

Directly using LLM to generate SQL often fails to yield executable queries. Therefore, providing similar cases as templates is an intuitive approach, similar to how users adjust their prompts when they fail to get correct code from the LLM.
Building on Baseline 1, we provide structured templates for tasks of varying difficulties to help the LLM generate more accurate SQL.
It is worth noting that this approach significantly increases the number of input tokens compared to Baseline 1.

\item Baseline 3: Using ReAct as agentic framework.

The ReAct (Reasoning + Acting) agent paradigm is currently a common approach to call tools for problem-solving, as demonstrated in \cite{zhang2024mapgpt,zhang2024geogpt}.
It is a framework that enables language models to solve problems by observing the environment, reasoning, and executing actions. To maintain clarity and controllability in reasoning and actions, ReAct typically allows only one action per step. 

In this baseline, the functions available to the ReAct framework include the \textit{Query Code Generation} within the \textit{Entity Retriever} module for data retrieval and the \textit{Geo-Calculation} in the \textit{Data Analyzer} module for spatial calculations.
Unlike the natural language inputs used in our \textit{Mission Planner} agent, the input variables for these two functions have a fixed format, as all reasoning tasks must be fully handled by ReAct. 
Since iterative reasoning is required in ReAct to complete the task, this framework consumes significantly more input and output tokens compared to other frameworks. 
Furthermore, as the complexity of the problem increases, the average number of required iterations also rises, leading to an even greater increase in token usage.
Explanations of the functions and the prompts used for this baseline can be found in Appendix \ref{sec:prompt_ReAct}.\\
\end{itemize}

Table \ref{tab:correct_assess} compares the accuracy of our proposed approach with the three baseline methods across tasks of increasing complexity, each consisting of 100 queries. 
The results indicate that our approach consistently outperforms all baselines, and the performance gap becomes more pronounced as task complexity rises. 
For instance, in the simplest scenario of two geo-entities without modifiers, our method achieves 96\% accuracy, exceeding Baseline 1 and 2 by 20-22 percentage points, and Baseline 3 by 55 percentage points. 
In more complex scenarios involving modifiers, our approach continues to surpass the baselines by margins of up to 40 percentage points, demonstrating its robustness and effectiveness in handling queries of increasing complexity.

Table \ref{tab:tokens} summarizes the average number of tokens used for tasks of varying difficulty levels across different baseline methods. It is observed that Baseline 3 consumes significantly more tokens than all other methods. Our approach uses more tokens than Baseline 1 but fewer than Baseline 2. Additionally, as task difficulty increases, token usage also rises.

Further, it can be observed from Table \ref{tab:correct_assess} that including the templates (Baseline 2) generally yields better performance compared to not using them (Baseline 1). The number of input tokens in Baseline 2 is higher than in Baseline 1, while the output token count remains similar.
As the complexity of the dataset increases, the accuracy of directly generating SQL tends to decrease, while the number of tokens used increases.
The primary reason is that as SQL length and complexity increase, it is more likely that structural issues arise, as also evidenced in \cite{hong2024next}. However, by providing the LLM with corresponding templates, we observe that on more complex queries, Baseline 2 achieves a more obvious accuracy improvement over Baseline 1.
Moreover, it is evident that SQL generation heavily relies on naming conventions. If there are inconsistencies, even with a given template, it is still highly likely to produce queries that cannot be executed.

\begin{table}[width=\linewidth,cols=2,pos=h]
\caption{Averaged Number of Iterations using ReAct}
\label{tab:react}
\begin{tabular*}{\tblwidth}{@{} LC@{} }
\toprule
\textbf{Dataset} & \textbf{\# of Iterations}\\
\midrule
(1) 2 geo-entities without modifiers & 4.72 \\
(2) 2 geo-entities with modifiers  & 3.97 \\
(3) 3 geo-entities with modifiers  & 5.91 \\
(4) 4 geo-entities with modifiers  & 8.00 \\
\bottomrule
\end{tabular*}
\end{table}

Baseline 3 employs the ReAct paradigm agent framework. However, since all reasoning tasks—including entity retrieval and spatial calculation—must be handled by a single agent, the required context length for this framework's agent becomes significantly longer. 
As seen in Table \ref{tab:react}, as the dataset complexity increases, the number of iterations in Baseline 3 increases. 
This causes the context length to increase drastically in later iterations. 
According to \cite{wang2024tdag}, such long contexts negatively impact the model's reasoning performance. Moreover, experimental results show that the probability of function call errors in the ReAct framework also increases with the number of iterations.

Consequently, the model encounters very long contexts during the later stages of reasoning. On one hand, this leads to token waste, and on the other, it results in low reasoning efficiency. Additionally, long reasoning chains introduce uncertainty in subsequent reasoning steps. For instance, errors in a lengthy reasoning chain may gradually accumulate with each step, eventually leading to unreliable outcomes \citep{liu-etal-2024-trustworthiness,yao2023react}.

ReAct is inherently designed for complex problems that cannot be easily broken down through process-oriented decomposition. In contrast, our task of geospatial data retrieval and spatial analysis is highly structured and can be effectively decomposed into smaller, manageable subtasks. As a result, ReAct is not the best fit for this scenario.

For a structured workflow like ours, using a multi-agent approach can effectively distribute the workload among different agents and prevent context interference between different tasks. This not only reduces token consumption but also improves overall performance.

\subsection{User Tests}
\label{sec:user_tests}

For a barrier-free geoportal, evaluating system performance is crucial, but equal emphasis should be placed on assessing user experience. Incorporating user feedback ensures a comprehensive understanding of how effectively the portal meets accessibility and usability needs.

User Experience Questionnaire (UEQ)~\citep{schrepp2017construction} is a standardized instrument designed to measure the user experience of interactive products. 
It is intended to capture users' subjective impressions and emotional reactions to a product. 
The UEQ comprises several dimensions, each assessed through multiple questions, including \textit{Attractiveness}, \textit{Perspicuity}, \textit{Efficiency}, \textit{Dependability}, \textit{Stimulation}, and \textit{Novelty}.
This questionnaire consists of 26 descriptive items using a 7-point Likert scale, such as "confusing/clear", "complicated/easy", each represented by a pair of opposing words, with the chart showing which side each item leans toward.
By integrating the questions to those six dimensions, the UEQ provides a comprehensive evaluation of the interactive product's strengths and weaknesses. 
This questionnaire is widely utilized in usability studies and user experience research \citep[e.g.,][]{rey2024exploring,rey2024assessing} to collect quantitative data on user satisfaction.

\begin{figure}[ht]
    \centering
    \includegraphics[width=\linewidth]{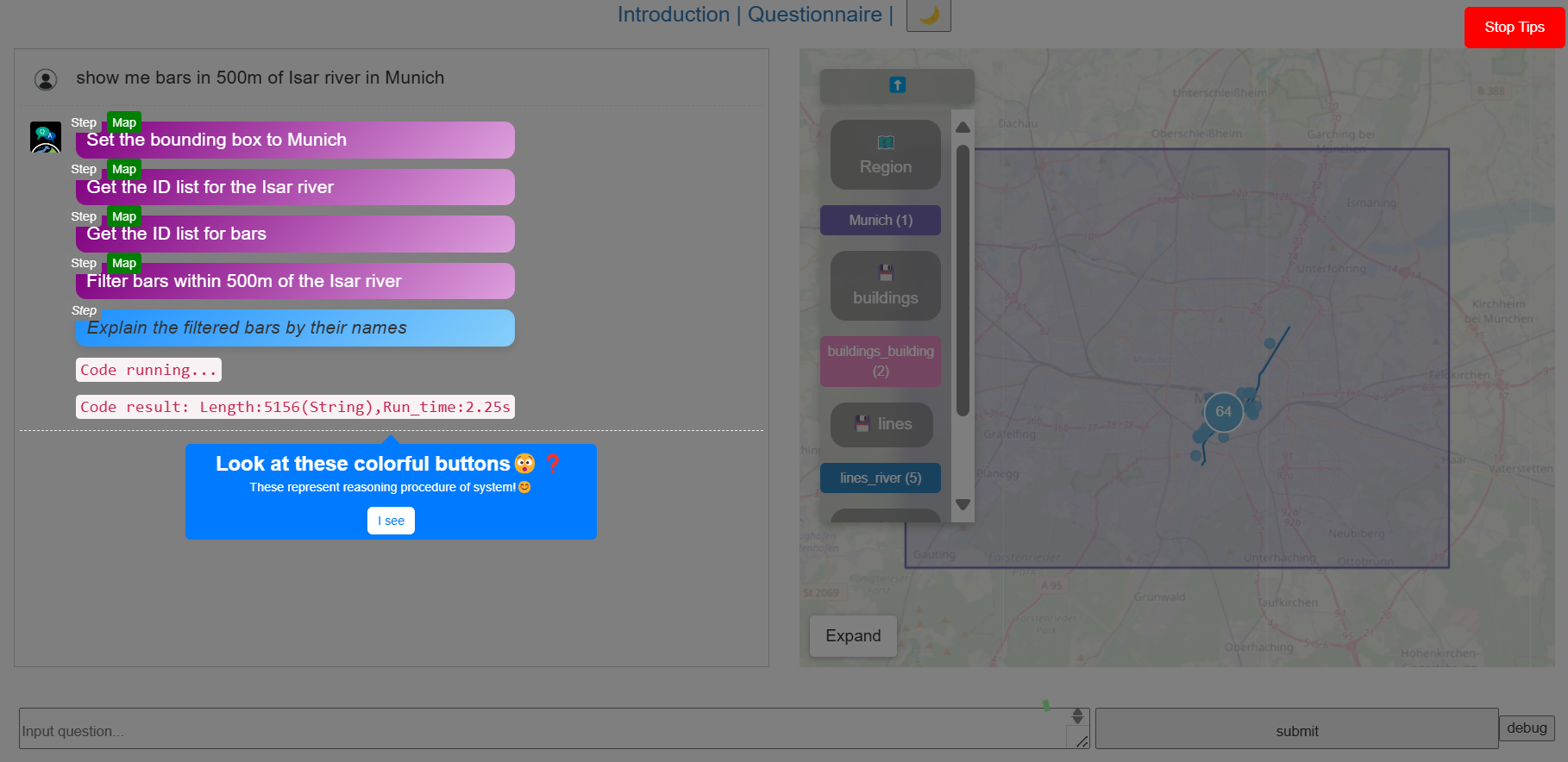}
    \caption{The tutorial interface highlighting areas for user interaction within the system.}
    \label{fig:tutorial}
\end{figure}

At the beginning of the study, participants received a link to access our system on their own laptops. They first went through a tutorial that included the four example questions as introduced in Section \ref{sec:case_studies}, as well as an introduction to the interface elements, highlighting areas where users can interact with the system, as shown in Figure \ref{fig:tutorial}. 
Following the tutorial, participants were free to explore the system at their own pace and engage in the chat to ask any questions, without time constraints. After completing their exploration, participants were asked to fill out a questionnaire. 
In addition to the standard UEQ questions, we included two additional questions regarding the participants' familiarity with GIS systems and SQL. This was done to assess how prior knowledge might affect their experience with the system. 
These questions were also rated on a 7-point Likert scale, with proficiency levels ranging from high (1) to low (7).

We conducted this user test among 85 voluntary participants from various backgrounds, including Computer Science, Land Surveying, Geographic Information Science, and Cartography. 
These fields covered primary potential users of a geoportal. 
Among the participants, 68 completed the entire tutorial before filling out the questionnaire, forming the basis of our follow-up analysis, while 17 were excluded since they did not complete the tutorial. 

\subsubsection{User experience analysis}

\begin{figure}[h]
    \centering
    \includegraphics[width=0.49\textwidth]{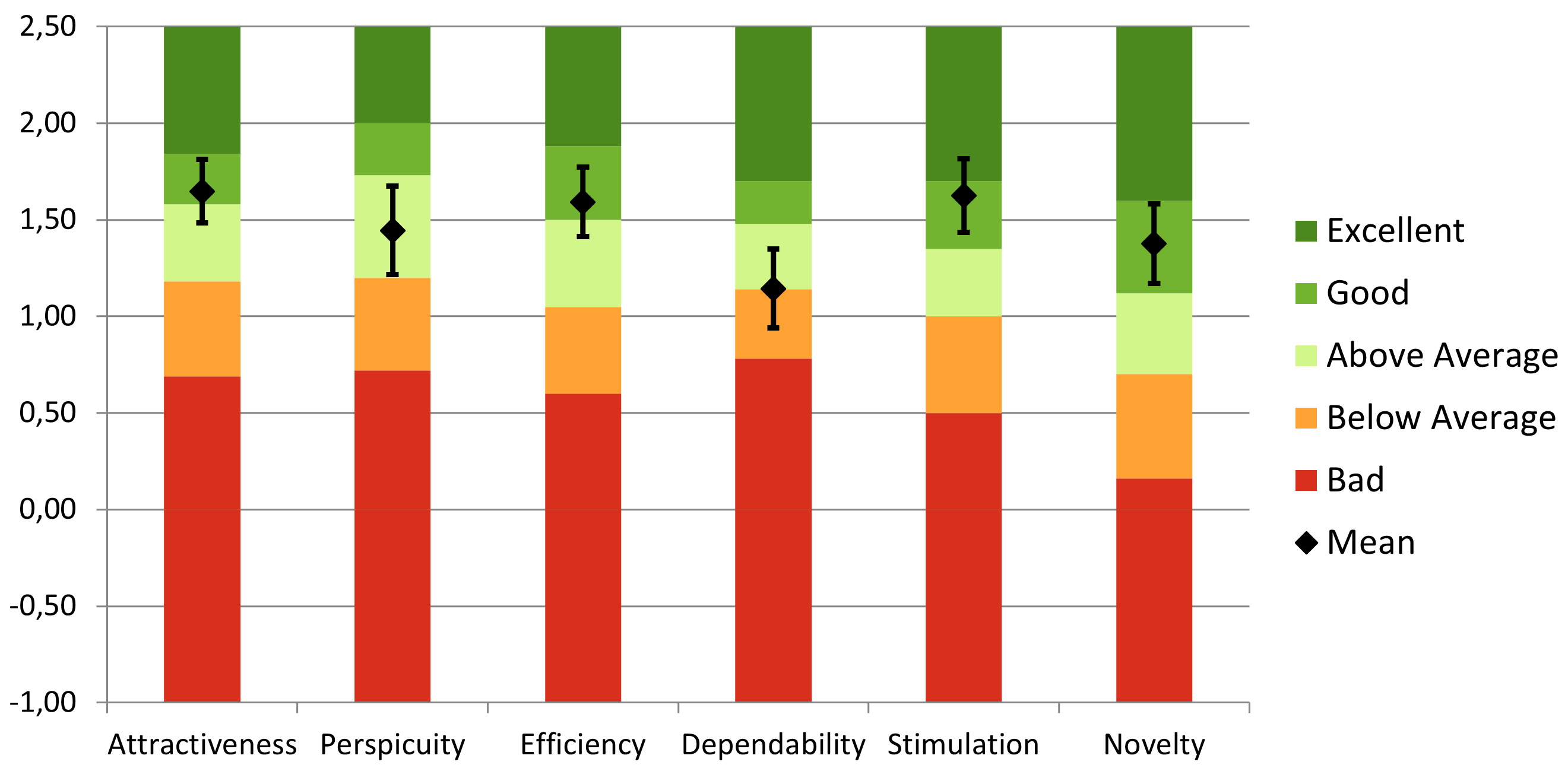}
    \caption{Scale means of the UEQ across all six scales with confidence intervals, for benchmarking relative product quality (N=68).}
\label{fig:UEQ}
\end{figure}

Figure \ref{fig:UEQ} presented the results of UEQ across the six experience measured scales. 
This can be compared with the official UEQ benchmark, which is based on the assessments from 452 products and 20,190 participants\footnote{User Experience Questionnaire Handbook (Version 11). Source: \url{https://www.ueq-online.org/Material/Handbook.pdf}}. 
It has been observed that for typical products evaluated so far, a sample size of approximately 20-30 participants tends to yield fairly stable results.
Compared to this benchmark \citep{schrepp2017construction}, the dimensions of \textit{Attractiveness}, \textit{Efficiency}, \textit{Stimulation}, and \textit{Novelty} fall within the \textit{Good} range, meaning that only 10\% of the benchmark data outperformed our system, while 75\% performed worse. The dimensions of \textit{Perspicuity}, and \textit{Dependability} fall within the \textit{Above Average} range, indicating that 25\% of the benchmark results were better than our system, while 50\% were worse.

\subsubsection{User knowledge in GIS and SQL}

\begin{figure}[h]
    \centering
    \includegraphics[width=0.5\textwidth]{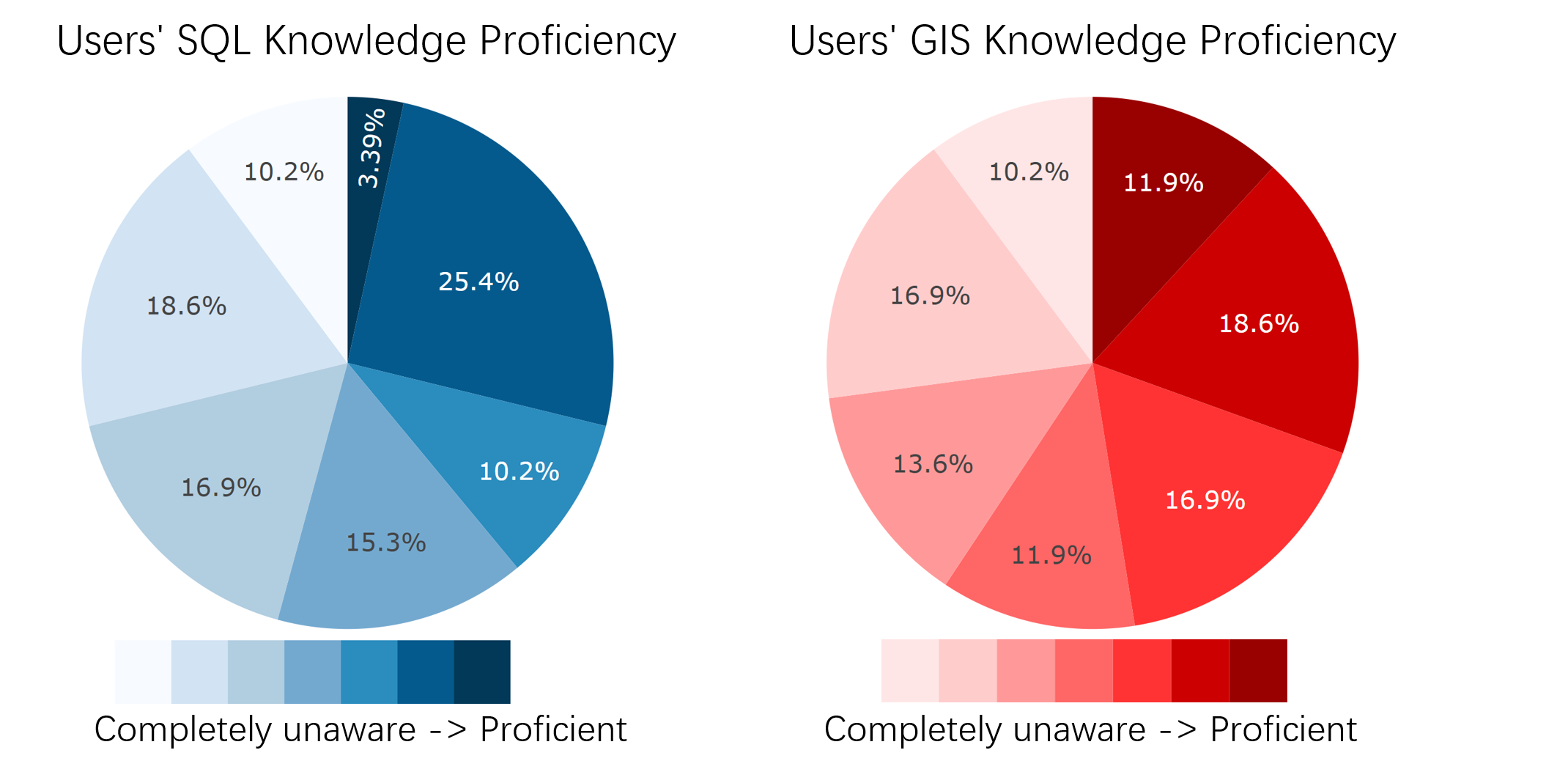}
    \caption{Participants' self-assessment of their familiarity with SQL query language (blue) and GIS System (red).}
\label{fig:familiar_gis}
\end{figure}

Out of the 68 participants, 59 provided self-assessments of their familiarity with SQL query language and GIS Systems. Their rating distribution is shown in in Figure \ref{fig:familiar_gis}. 
Participants' self-assessments of their GIS and SQL skills were relatively evenly distributed.
The number of participants who rated themselves as relatively proficient (selecting 1 or 2) was roughly equal to those who considered themselves completely unaware (selecting 7 or 8).
Therefore, our user test effectively reflects the preferences and needs of both proficient and less proficient users, ensuring the design accommodates this diverse audience.

\begin{figure}[h]
    \centering
    \includegraphics[width=0.5\textwidth]{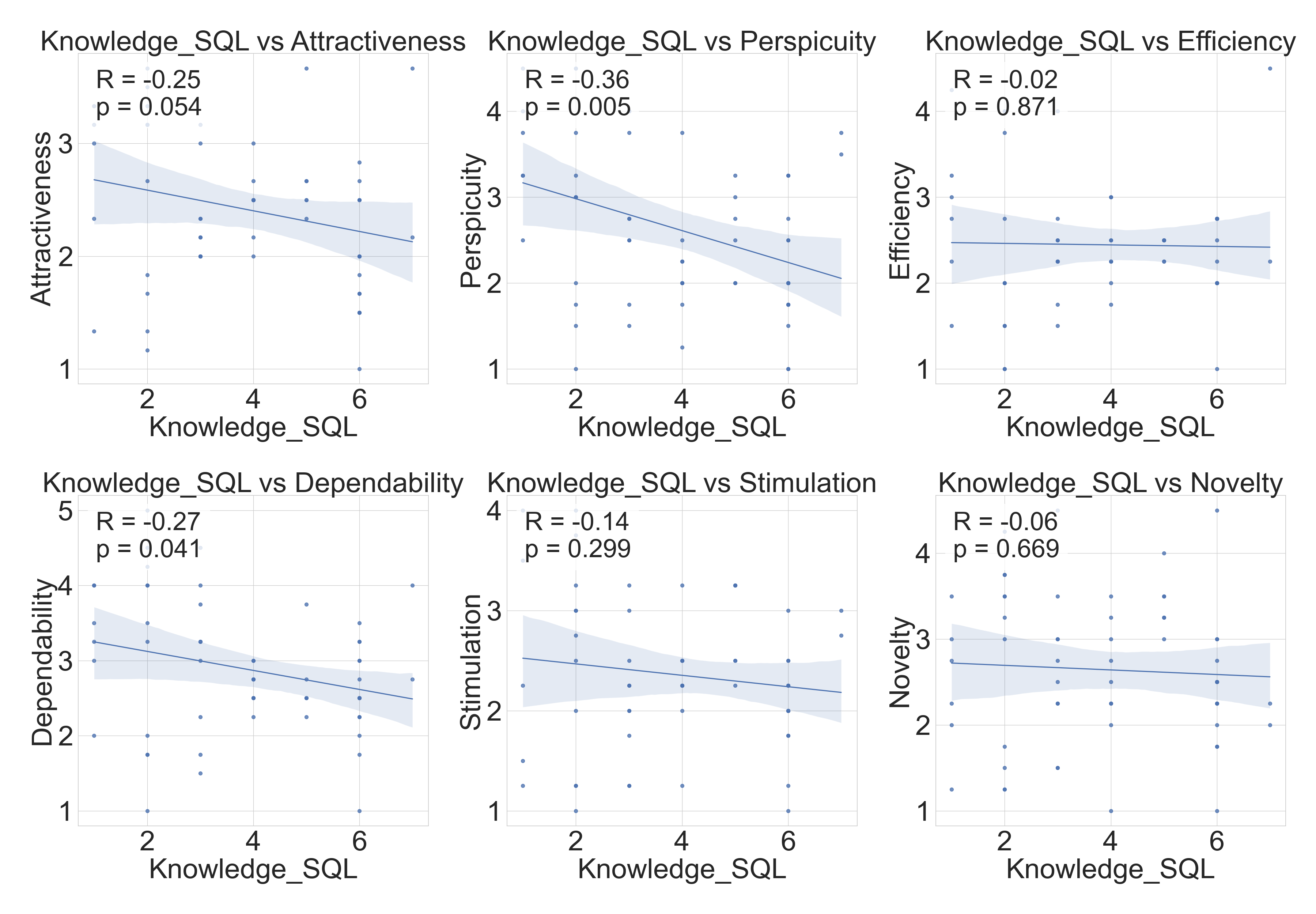}
    \caption{Correlation analysis between SQL knowledge and user experience of six UEQ scales.}
\label{fig:SQLCorrelation}
    \centering
    \includegraphics[width=0.5\textwidth]{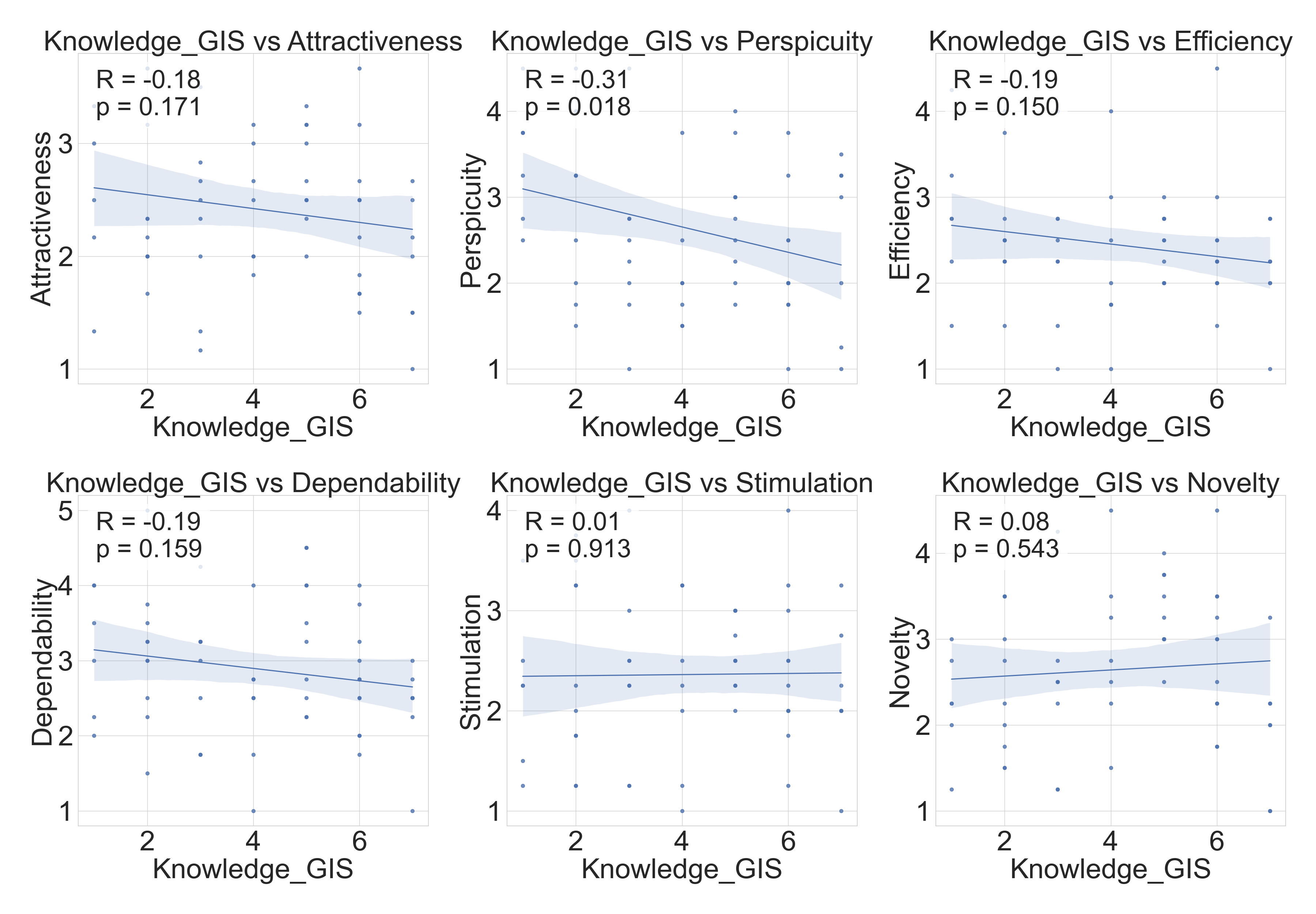}
    \caption{Correlation analysis between GIS knowledge and user experience of six UEQ scales.}
\label{fig:GISCorrelation}
\end{figure}

Next, we aim to investigate whether users with different skill levels and proficiencies show variations across the six scales measured in our study.
Figure \ref{fig:SQLCorrelation} and \ref{fig:GISCorrelation} show the relationship between users' knowledge in SQL and GIS and their evaluation of the system in terms of the six scales of UEQ. 
The horizontal axis in the figures represents the level of knowledge, where a higher value indicates lower proficiency. 
A negative correlation (\( R < 0 \)) suggests that users with higher knowledge tend to rate the attribute more positively, with \( p < 0.05 \) considered statistically significant.

For SQL query language, significant negative correlations are found for \textit{Perspicuity} ($R = -0.36, p = 0.005$), and \textit{Dependability} ($R = -0.27, p = 0.041$), indicating that users with higher SQL knowledge rated these aspects more positively. Although \textit{Attractiveness} and \textit{Efficiency} showed a negative trend, it was not statistically significant. 

Similarly, for GIS systems, significant negative correlations were found for \textit{Perspicuity} ($R = -0.31, p = 0.018$), suggesting that users with higher GIS knowledge gave more favorable ratings on this scale. No other significant correlations were observed for GIS knowledge.

This suggests that higher proficiency in both SQL and GIS generally leads to more positive perceptions of the system’s \textit{Perspicuity} and \textit{Dependability}. 
Other aspects such as \textit{Attractiveness}, \textit{Efficiency}, \textit{Stimulation} and \textit{Novelty} are not significantly influenced by users' knowledge levels. Nevertheless, these scales are generally evaluated positively on average.

\subsubsection{Comment analysis}

\begin{table*}[ht]
\centering
\caption{User Comments}
\label{tab:user_comments}
\begin{adjustbox}{max width=\textwidth}
\begin{tabular}{cp{12.5cm}}
\toprule
\textbf{Index} & \textbf{Comments} \\
\midrule
 &  \textbf{Praise}\\
1    &  A great innovative concept, with much consideration and integration with other Open Geographic information service providers, it can be one of the best and greatest innovations of the moment. \\
2    &  Nice and motivating work! \\
3    &  Do you have business plan? \\
4    &  pretty easy to handle, nice Ui, but tutorial maybe too long, I mean we need a pause function \\
5     &  Better with route ability like Google, I think your app would be more feasible in route than google in some way...like spatial relationship, I want to search accurate result like 'shopping mall in 1000m of highway', google cannot give me. Good anyway \\
\\
 &  \textbf{Performance Issue}\\
6     &  It could answer regular questions. Also, it had a problem finding cinemas. \\
7     &  Sometimes entering prompts won't give the expected results... \\
8    &  I love your idea, but sometimes it is too slow, maybe network \\
9     &  the messages pop up to the tutorial were annoying which I could not be able to use the app \\
10     &  I can't scroll through all the different filters in the map on the left. It has the region, then all the points but one can't scroll down to select/deselect them \\
\\
 &  \textbf{Suggestions}\\
11     &  Better describe what kind of problems could be solved \\
12    &  Is it possible to open my own data? \\
13     &  The chart generated is larger than the chat box which is a little bit annoying, I try to adjust it but it didn't perform as expected. (maybe because I raise new questions and it can't remember the previous ones? or it don't have this kind of functions?) \\
14     &  It would be better if the map could take a larger space, although it could be expanded, it's more convenient for us to view map and the chat box at the same time. (me personally) \\

\bottomrule
\end{tabular}
\end{adjustbox}
\end{table*}

Table \ref{tab:user_comments} shows all 14 of the user's comments on our system, where we grouped them into 3 categories.
Overall, most of the issues raised by users are concentrated on the user interface (UI) and interaction aspects, such as network latency (the delay in response might be due to the high number of users interacting with the system simultaneously) and interaction design flaws within the UI. 
With these suggestions, we adapted the UI and the system tutorial accordingly to meet the user's expectations. 
Despite these concerns, the feedback strongly recognizes the innovative nature of the system and considers it an interesting product in the GIS field.

\section{Discussion}

In this work, we present our proposed system, which demonstrates superior performance compared to several baseline methods (including SQL generation and ReAct) and delivers an exceptional user experience, as confirmed through user testing.
However, there are still many limitations for the system, that would need future development. We will discuss the challenges and opportunities from the following three aspects:

\subsection{Uncertainty of Users' Expression}

Users' queries are often vague or ambiguous, such as asking for "nearby buildings". "nearby" can mean buildings that touch the park boundary, or those within a certain distance from it, leading to different interpretations. 
Additionally, precise geographic relationships can often be counterintuitive. 
For instance, the distinction between "intersects" and "contains" is frequently confusing for users. Intuitively, users may describe a scenario where one entity mostly contains another but overlaps slightly as "contains." However, in precise geographic terminology, such a relationship is classified as "intersects" due to the overlap. 
This discrepancy between intuitive language and precise geographic terminology creates challenges for the system in accurately interpreting user intent in spatial relationships.

It is necessary for the system to ask the user for clarification when uncertain interpretations arise in order to better determine the user's intent. 
However, the ambiguity in the formulation of spatial queries still requires further research to comprehensively identify and address potential ambiguities.


\subsection{Impact of Formatted Outputs on LLM Performance}

In our framework, due to programmatic requirements, agents often return structured outputs such as booleans or JSON. However, as observed and validated in \cite{tam2024letspeakfreelystudy}, formatted outputs can negatively impact LLM reasoning due to their lack of inherent reasoning flexibility. To address this, we require agents to first produce a chain-of-thought output \citep{wei2023chainofthought} in all cases, where structured results are needed. Once the reasoning process is outlined, the final structured output should be encapsulated in the specified format, ensuring that programs can extract the results seamlessly.
Despite this approach, our observations suggest that even this method may still yield suboptimal performance compared to unconstrained outputs. 

\subsection{Optimization of query process}

Typically, databases like PostgreSQL generate multiple possible plans during query execution. The query optimizer evaluates each potential execution plan and selects the one expected to execute the fastest\footnote{Chapter 50. Overview of PostgreSQL Internals. 50.5. Planner/Optimizer. Source: \url{https://www.postgresql.org/docs/current/planner-optimizer.html}}. For spatial tasks in PostGIS, a similar approach leverages spatial indexes and cost-based analysis to optimize query performance. However, in our system, geographic information is retrieved from each respective database into memory and processed using the Shapely library, which lacks native optimization. This method, while necessary to support cross-database queries, reduces execution efficiency—even with prior spatial filtering performed in the databases. Looking ahead, a potential optimization direction could involve integrating native database optimizations with in-memory processing to enhance the overall performance.

\section{Conclusion and Outlook}

In this work, we present an innovative framework for geospatial data retrieval and analysis with a barrier-free geoportal that allows users to interact with geospatial data through natural language. In particular, users without GIS or database expertise can effortlessly engage with geospatial data.
This framework incorporates several novel ideas, including:
\begin{itemize}
    \item Leveraging a vector database based on word embeddings to achieve semantic-based data retrieval, ensuring more accurate search results while allowing for a certain degree of fuzziness.
    \item Adopting a multi-agent LLM system, with each agent thoughtfully designed to focus on a single task, enhancing robustness and the success rate of complex geospatial analyses.
    \item Building a cross-database knowledge graph for cross-database intersectional searches, allowing data from different databases to be seamlessly integrated and queried together.
    \item Allowing users to explore geospatial data entirely through natural language, without the need for direct interaction with raw geospatial data.
    \item Visualizing analytical steps to promote user understanding, enhance process transparency, and build trust, particularly for non-expert users.
\end{itemize}

The performance of the system has been validated through a series of case studies and evaluated both systematically and through user testing.
The evaluation results indicated that our framework outperforms baseline methods, including SQL generation with and without templates, and ReAct, in terms of accuracy, robustness, and usability. Notably, the framework excelled in handling complex geospatial analytical queries.  Furthermore, in certain scenarios, the framework demonstrated the ability to optimize the query process by reducing the number of tokens required, resulting in more efficient processing.
User testing demonstrated the potential for intuitive geospatial data interaction, allowing users without extensive technical knowledge to effectively perform complex queries. The feedback showed strong user satisfaction, with positive usability and accessibility, while identifying areas for improvement in the user interface and system responsiveness.

These results underscore the potential of the proposed framework to bridge the gap between non-expert users and geospatial data, offering a practical solution for broader accessibility and usability. 
Future work will focus on further improving the system's scalability, extending its support for more complex geospatial analysis, and integrating real-time user feedback to refine performance.

\section*{Declaration of competing interest}
No conflict of interest exists in the submission of this manuscript, and the manuscript is approved by all authors for publication. The work described was original research that has not been published previously, and not under consideration for publication elsewhere, in whole or in part.

\section*{Acknowledgement}
The authors would like to acknowledge the support from the German Research Foundation (DFG) funded research project ``Dense and Deep Geographic Virtual Knowledge Graphs for Visual Analysis'' (DFG, 500249124) and NFDI4Earth - NFDI Consortium Earth System Sciences (DFG, 460036893).

\appendix

\section{Prompt of Router}
\label{sec:prompt_router}

\begin{MyVerbatim}
You are a router responsible for directing incoming prompts to either the Analyzer or the Explainer based on the user's request. 

1. If the user's request involves new calculations or queries related to geographical entities, forward it to the Analyzer. For example:

   **Prompt**: I want to know which buildings are within 100m of the forest. 
   **Response**:  
   ```json
   {
       "Receiver": "Analyzer",
   }
   ```

2. If the user's request involves analyzing or displaying the results of a previous query, such as explanation, result interpretation, or visualization, and does not require new spatial calculations, forward it to the Explainer. For example:

   **Prompt**: I want to know what data I have.  
   **Response**:  
   ```json
   {
       "Receiver": "Explainer",
   }
   ```

Act accordingly for any given user prompt. Response in Json format.
\end{MyVerbatim}

\section{Prompt of Relation Analyzer}
\label{sec:prompt_rel_analyzer}

\begin{MyVerbatim}
You are an excellent linguist, Help me identify all entities from this statement and spatial_relations. Please format your response in JSON. 
Reasoning before giving results.
Example:
query: "I want to know which soil types the commercial buildings near farm in Munich"
response:
{
"entities":
[
  {
    'entity_text': 'soil',
  },
  {
    'entity_text': 'commercial buildings',
  },
    {
    'entity_text': 'farm',
  }
],
 "spatial_relations": [
    {"type": "on", "head": 1, "tail": 0},
    {"type": "near", "head": 1, "tail": 2}
  ],
  "region": "Munich"
}

query: "I want to know residential area in around 100m of land which is forest in Maxvorstadt"
response:
{
  "entities": [
    {
      "entity_text": "residential area",
    },
    {
      "entity_text": "land which is forest",
    },
  ],
  "spatial_relations": [
    {"type": "in around 100m of", "head": 0, "tail": 1},
  ],
  "region": "Maxvorstadt"
}
query: "show land which is university and has name TUM in Munich Maxvorstadt"
response:
{
  "entities": [
    {
      "entity_text": "land which is university and has name TUM",
    },
  ],
  "spatial_relations": [],
  "region": "Munich Maxvorstadt"
}
query: "show land which is university or bus stop"
response:
{
  "entities": [
    {
      "entity_text": "land which is university or bus stop",
    },
  ],
  "spatial_relations": [],
  "region": ""
}
Notice, have/has should be considered as spatial_relations:
like: residential area which has buildings.

\end{MyVerbatim}

\section{Prompt of Mission Planner}
\label{sec:prompt_functions}

\begin{MyVerbatim}
You have following tools available to answer user queries, please only write python code:

1.set_bounding_box(address):
Input:An address which you want search limited in.
Output:None, it establishes a global setting that restricts future searches to the defined region.
Usage:By providing an address, you can limit the scope of subsequent searches to a specific area. This function does not produce any output, but it establishes a global setting that restricts future searches to the defined region. For example, if you want to find buildings in Munich, you should first set the bounding box to Munich by using set_bounding_box("Munich").
Notice:Please include the directional words like east/south/east/north of query in the address sent to set_bounding_box
Notice:If user does not query in a specific area, do not use this function. If user wants to search in all area, call set_bounding_box('').

2.id_list_of_entity(description of entity):
Input: Description of the entity, including adj or prepositional phrase like good for commercial,good for planting potatoes, or just entity word like 'Technische Universität München'. Make sure to make plural words singular when you input.
Output: A list of IDs (id_list) corresponding to the described entity.
Usage: Use this function to obtain an id_list which will be used as input in the following functions.
Notice: Some times the description may have complex description like:"I want to know land which named see and is water", input the whole description into function.
Notice: Do not input geographical relation like 'in/on/under/in 200m of/close' into this function, it is not description of entity.

3.geo_filter('their geo_relation',id_list_subject, id_list_object):
Input: Two id_lists (one as subject and one as object) and their corresponding geographical relationship.
Output: A dict contains 'subject','object' two keys as filtered id_lists based on the geographical relationship.
Usage: This function is used only when the user wants to query multiple entities that are geographically related. Common geographical relationships are like: 'in/on/under/in 200m of/close/contains...'
Notice: id_list_subject should be the subject of the geo_relation, in example: soil under the buildings, soil is subject; buildings around water, buildings are subject.
Notice: Get the filtered subject/object id_list: result['subject'],result['object']
Reasoning before giving code.
Please always set an output variable for each function you called and write corresponding short code comments.
Variable in history is available to call.
\end{MyVerbatim}









\section{Prompt of Explain Agent}
\label{sec:prompt_explain}
\begin{MyVerbatim}
You need to write code to answer user questions, such as drawing charts (python) and handling database-related queries (Cypher).  

### Chart Drawing:  
- If the user asks for a diagram, always use the true variables from the previous code rather than assuming fake values.  
- If multiple graphs need to be drawn simultaneously, use subplots to display them in a single figure.  

### Database Queries:  
- Write Cypher code to read a graph to answer database-related questions.  
- The graph contains information about a database where each **column value** and **table** is stored as a node, and relationships between them are stored as links.  

#### Graph Structure:  
- **Nodes (`nodes`)** have two attributes:  
  - `type`: The type of the node (e.g., `table`, `name`, `fclass`).  
  - `id`: The specific identifier of the node.  
- **Edges (`links`)** have three attributes:  
  - `edge_type`: A string separated by `_` indicating relationships.  
    - Example:  
      - `table_fclass`: An edge from a `table` node to an `fclass` node.  
      - `table_fclass_reverse`: An edge from an `fclass` node to a `table` node.  
  - `source`: The source node.  
  - `target`: The target node.  

### Cypher Code Requirements:  

- The cypher code should always start with ` ```cypher` and end with ` ``` `.  
- If the code executes correctly, the system will return results, and you must include `"Explain result"` in your response.  
\end{MyVerbatim}
\section{Prompt of Intent Matcher Agent}
\begin{MyVerbatim}
You will be provided with a geographic information-related query, along with supporting details, and your task is to analyze and respond based on the following guidelines:

Input Details:
A query related to geographic information (e.g., "Chinese restaurants," "Isar River").
A list of table names from the current database.
A partially matched result linking the query to database information.
Tasks:
Table Association: Determine if the query is strongly related to a specific table.
If strongly related, return the table name under the key "table".
If no strong relation exists, return an empty string ("").
Match Relevance: Evaluate whether the partially matched result aligns with the query's intent.
For example, if the query is "Chinese restaurants" and the match is "tourist info" restaurants, it’s not suitable.
Return a list of appropriate matches under the key "valid_pairs".
Query Intent: Identify whether the query is based on a name or a category.
Name-based examples: "Isar River," "Hauptbahnhof" (return "named_entity": true).
Category-based examples: "greenery space," "educational institute" (return "named_entity": false).
Output Format: Return your response as a JSON object.
Reasoning before giving result.
```json
{
  "named_entity": true/false,
  "valid_pairs": [],
  "table": ""
}
```
\end{MyVerbatim}
\section{Prompt of ReAct Agent}
\label{sec:prompt_ReAct}
\begin{MyVerbatim}
    
### ** ReAct Agent Prompt: Thought → Action → Observation**  

#### **Task:**  
Develop an agent that follows the **ReAct framework** (Thought → Action → Observation) using the tools below:  

---  

### ** Available Tools**  

1. **`get_ids()`**  
   - Input: `{ "type": "", "name":""}`  
   - Returns a list of elements based on type/name.  
   - If it returns 0, you need to reconsider the input and search again. 

2. **`geo_calculate(id_list_subject, id_list_object, geo_relation_type, buffer_num)`**  
   - Computes spatial relationships: `["intersects", "buffer", "in", "contains"]`.  
   - Outputs `{ "subject": [...], "object": [...] }`.  

---

### ** Process**  
1. **Thought:** Determine required elements, relationships, and buffer size.  
2. **Action:**  
   - Call `get_ids()` to retrieve elements.  
   - Use `geo_calculate()` for spatial computations.  
3. **Observation:** Evaluate results. Iterate if needed.  

---

### **Rules**  
- Use **only one action per step**.  
- **Store function outputs in variables** for reuse.  
- Assign the final id list to `final_result`, if it is empty, assign it to zero: final_result=0

**Example Format:**  
```python
ids_subject = get_ids({ "type": "", "name": "" })
```

Now, **begin!**
\end{MyVerbatim}



\printcredits

\bibliographystyle{cas-model2-names}

\bibliography{cas-refs}


\end{document}